\documentclass[onecolumn,12pt,journal,draftclsnofoot]{IEEEtran}
\usepackage[latin9]{inputenc}
\usepackage{color}
\usepackage{textcomp}
\usepackage{amsmath}
\usepackage{amssymb}
\usepackage{graphicx}
\PassOptionsToPackage{normalem}{ulem}
\usepackage{ulem}
\usepackage{colortbl}
\definecolor{mygray}{rgb}{.83, .83, .83}
\definecolor{mygreen}{rgb}{0.9608,    0.9608 ,   0.8627}
\definecolor{mycyan}{rgb}{0.9412,    1.0000,    1.0000}

\makeatletter

\providecommand{\tabularnewline}{\\}

\usepackage{amsfonts}
\usepackage{mathrsfs}
\usepackage{mathrsfs}\usepackage[noblocks]{authblk}
\usepackage[compress]{cite}
\usepackage{bm}
\usepackage{algorithm}
\usepackage{algorithmic}
\usepackage{epsfig}\usepackage{graphics}\usepackage{subfigure}\usepackage{epsfig}\usepackage{epstopdf}
\usepackage{graphics}\usepackage{subfigure}\usepackage{upgreek}\usepackage{caption}\usepackage{theorem}
\usepackage{breqn}
\usepackage{multicol}
\usepackage{eufrak}
\usepackage{eucal}
\usepackage{array}
\usepackage{cases}
\usepackage[compress]{cite}
\usepackage{algorithm}
\usepackage{algorithmic}
\usepackage{empheq}
\allowdisplaybreaks[4]

\newtheorem{lemma}{Lemma}

\theoremheaderfont{\normalfont\bfseries}

\makeatother

\begin{document}
\title{Radar Rainbow Beams For Wideband mmWave Communication: Beam Training And Tracking}
\author{Gui~Zhou, Moritz Garkisch, Zhendong Peng, Cunhua~Pan, Robert Schober 
 \thanks{G. Zhou, M. Garkisch, and R. Schober are with the Institute for Digital Communications,
Friedrich-Alexander-University Erlangen-Nürnberg (FAU), 91054 Erlangen,
Germany (email: gui.zhou, moritz.garkisch, robert.schober@fau.de). 
Z. Peng is with the Department
of Electrical and Computer Engineering, The University of British
Columbia, Vancouver, BC V6T 1Z4, Canada. (zhendongpeng@ece.ubc.ca).
C. Pan is with the
National Mobile Communications Research Laboratory, Southeast University,
Nanjing 210096, China. (cpan@seu.edu.cn). }}
\maketitle

\begin{abstract}
	We propose a novel integrated sensing and communication (ISAC) system  that leverages sensing to assist communication,  ensuring fast initial access, seamless user tracking, and uninterrupted communication for millimeter wave (mmWave) wideband systems. True-time-delayers (TTDs) are utilized to generate frequency-dependent radar rainbow beams by controlling the beam squint effect. These beams cover users across the entire angular space simultaneously for fast beam training using just one orthogonal frequency-division multiplexing (OFDM) symbol.  Three detection and estimation schemes are proposed based on radar rainbow beams for estimation of the users'  angles, distances, and velocities, which are then exploited for  communication  beamformer design.  The first proposed scheme utilizes a single-antenna radar receiver and one set of rainbow beams,  but may cause a Doppler ambiguity. To tackle this limitation, two additional schemes are introduced, utilizing two sets of rainbow beams and a multi-antenna receiver, respectively.    Furthermore, the proposed detection and estimation schemes are  extended to realize  user tracking  by choosing different subsets of OFDM subcarriers. This approach eliminates the need to switch phase shifters and TTDs, which are typically necessary in existing tracking technologies, thereby reducing the demands on the control circurity.
Simulation results reveal the effectiveness of the proposed rainbow beam-based  training and tracking methods for mobile users.  Notably, the scheme employing a multi-antenna radar receiver can accurately estimate the  channel parameters and can support communication rates comparable to those achieved with perfect channel information.

\end{abstract}


\section{Introduction}

The integration of sensing and communication (ISAC) has garnered substantial
research interest due to the shared characteristics between sensing
and communication systems, including signal processing algorithms,
 system architecture, and spectrum \cite{2011radarproc,FanLiu-2022JSAC,Andrew-2021JSTSP}.
Sensing targets can be categorized into two groups: cooperative
targets with communication needs and non-cooperative targets, including
temporarily non-communicating users and obstacles. Sensing 
can be employed to detect and track both categories of targets. Utilizing
bidirectional radar signals for cooperative user detection and location
awareness offers advantages,  such as enabling pilot- and feedback-free, low-latency
communication, over unidirectional training communication signals. Sensing
is essential for detecting non-cooperative targets and constructing
a channel knowledge map \cite{zeng-2021Mag}. Such a map is valuable
for predicting communication disruptions based on the positions of
obstacles and user mobility patterns. It enables proactive selection
of alternative communication links, such as non-line-of-sight (NLOS)
paths reflected by environmental clutter or reconfigurable intelligent surface (RIS)/relay-aided links, thus
supporting resilient communication.

In both communication and radar systems,  beam training is of fundamental importance and requires probing beams pointing in different directions sequentially to detect the presence and estimate the positions of users/targets \cite{basic-training}. However, the linear scaling of the required number of beams with the array size gives rise to a large beam training overhead and latency. To enhance training efficiency, two common approaches are employed. The first approach, based on advanced signal processing, is capable of reducing the required number of beams logarithmically with respect to the array size. Methods like hierarchical beam sweeping \cite{Hierarchical-Codebook} and side-information-assisted beam training \cite{training-sideinformation} fall into this category. The second approach is based on multi-beam solutions, relying on advanced hardware design \cite{2014Asilomar,hardware-multibeam-2,hardware-multibeam,rainbow-Mag,2019Asilomar,rainbow-TCS2021}, to enable simultaneous channel probing. One method in this category is the simultaneous utilization of multiple radio frequency (RF) chains to generate beams pointing in different directions, where different beams scan different non-overlapping sectors \cite{2014Asilomar,hardware-multibeam-2,hardware-multibeam}. However, this approach incurs high hardware complexity and cost as a large number of RF chains is needed to reduce training time. Alternatively,  a cost-effective method is to employ simple true-time-delayers (TTDs), enabling the generation of frequency-dependent beams pointing in different directions by adjusting the TTD delays  \cite{2019Asilomar,rainbow-Mag,rainbow-TCS2021}. This facilitates the creation of multiple frequency-dependent beams covering the entire angular space by utilizing different frequencies. These so-called  rainbow beams can be readily generated based on multi-carrier waveforms  in millimeter wave (mmWave) systems.

The concept of TTD based rainbow beams, covering the full angular space, was initially introduced in \cite{2019Asilomar} for fast initial access in mmWave systems. Compared with single-carrier beam scanning exploiting the same bandwidth, multi-carrier rainbow beam training  may not have a big advantage in terms of training time. However, it does not need  to reconfigure  phase shifters (PSs) and TTDs  \cite{rainbow-Mag}.  This feature greatly reduces the requirements on the control circuits for beam training in terms of accuracy and switching time, making TTD based rainbow beams  an appealing approach, especially in mmWave systems, where sophisticated hardware based solutions are difficult to realize. 

 There has been some preliminary research on rainbow beam technology. The authors of \cite{rainbow-TCS2021} investigated  two TTD architecture candidates, namely analog and hybrid analog-digital
arrays, for rainbow beam generation. Building on these initial findings,
the authors of \cite{rainbow-linglong-tracking} extended this technique for user
tracking by generating rainbow beams covering  a narrow angular range to encompass possible user movements over a short duration.  Moreover, by leveraging the functional relationship between the user channel characteristics and user angles and distances in the near field, the authors of \cite{rainbow-linglong-nearfield} and \cite{rainbow-feifei-nearfield} explored near-field rainbow beam technology for fast training and localization, respectively.  Recently, rainbow beam technology has also found application in radar systems for the detection of the angles and distances of static targets \cite{rainbow-feifei-radar}.
Overall, rainbow beam technology has been shown to be effective in determining user/target positions, including angle and distance, in both communication and radar systems. However, the existing work in this area has two primary limitations: 1) The existing literature has primarily  focused on static targets and has not  considered  moving targets and velocity parameter estimation.  2)  Prior work has concentrated on the sensing
function of rainbow beams, either for detection and tracking of user position parameters using communication signals \cite{2019Asilomar,rainbow-TCS2021,rainbow-linglong-tracking,rainbow-linglong-nearfield,rainbow-feifei-nearfield} or detection of target position parameters using radar signals \cite{rainbow-feifei-radar}.
The application of rainbow beam technology in ISAC systems has not been explored, yet.

Against the above background, this work proposes a novel ISAC architecture aimed at enhancing communication through integration of radar functionality. The considered dual-function radar and communication (DFRC) base station (BS) employs a single RF chain to transmit sensing signals for user detection, estimation, and movement tracking. Leveraging timely and reliable channel information acquired through sensing, the DFRC BS further uses  multiple RF chains for communication with multiple users.  Considering a mmWave wideband system featuring massive antenna arrays,  TTDs  are utilized to respectively suppress the beam squint effect  for effective communication and harness the beam squint effect for generating  massive beams concurrently for sensing.   The major contributions of this work can be summerized as follows:
\begin{itemize}
\item First, we propose a radar multi-beam transmission scheme based on TTDs and orthogonal frequency-division multiplexing (OFDM) signals. Considering the requirements on angular coverage and the resolution of multiple beams, we derive a mathematical relationship between the required number of subcarriers and the antenna array size to create overlapping rainbow  beams. We also establish that the angular resolution of rainbow beams is unaffected by the subcarrier spacing and bandwidth.

\item Subsequently, we model the  radar rainbow beam echo signal and propose three corresponding detection and estimation schemes. The first scheme employs a single-antenna receiver to capture the echo signal of  one set of rainbow beams. It exploits the discrete Fourier transform (DFT) and a one-dimensional search to estimate the central reflection subcarrier frequencies and the corresponding angles of  the codebook for the users. The remaining user parameters, including velocity and distance, are successively estimated combining the  echoes of the subcarriers adjacent to the central reflection subcarrier.  While this scheme minimizes array and signal resource usage, it faces challenges in terms of distance and velocity estimation ambiguities. To address the velocity estimation ambiguity, we introduce a second scheme, which is based on the transmission of two sets of radar rainbow beams and leverages  the phase difference between two adjacent OFDM symbols to estimate high velocities, and a third scheme, which employs a multi-antenna receiver to capture the echo of one set of rainbow beams. 

\item Following this, based on the designed rainbow beams, an efficient tracking scheme is developed by manipulating a selected subset of subcarriers. The resulting codebook-based user tracking does not require the switching of PSs and TTDs, eliminating the need of rapid hardware reconfiguration  and the associated time overhead. Moreover, by allocating different frequency bands for tracking and communication, mutual interference between sensing and communication is effectively avoided,  facilitating continuous user tracking while maintaining uninterrupted communication.

\item Through simulations,  we show that,  for a single-antenna receiver, the proposed rainbow beam training scheme achieves comparable estimation performance as single-carrier radar systems.  Furthermore,  our result reveal that the benefits of rainbow beams can be fully leveraged by  multi-antenna radar receivers, yielding ultra-high precision for estimation. A 3 dB-beamwidth overlap is shown to be sufficient for rainbow beams to achieve high estimation performance. Moreover, our simulations confirm that all three estimation schemes proposed  can effectively track high-velocity, non-linearly moving users, acquiring  essential channel information for communication.  Notably, the scheme based on a multi-antenna radar receiver offers highly accurate angle information, resulting in nearly ideal communication rates for mobile  users. 
\end{itemize}
The remainder of this paper is organized as follows. In Section II, we introduce the system model and discuss  the frame structure of the considered ISAC system. In Section III, the design of the proposed rainbow beams is provided. The proposed rainbow  beam assisted fast beam training and user tracking schemes are  presented in Section IV. Finally, Sections V and VI report our numerical results and conclusions,
respectively.

\noindent \textbf{Notations:} The following mathematical notations and symbols are used throughout this paper. Vectors and matrices are
denoted by boldface lowercase letters and boldface uppercase letters, respectively. $\mathbf{X}^{*}$, $\mathbf{X}^{\mathrm{T}}$, $\mathbf{X}^{\mathrm{H}}$,
$\mathbf{X}^{\dagger}$, $||\mathbf{X}||_{F}$, $\mathrm{Tr}\{{\bf X\}}$, and $\mathrm{vec}(\mathbf{X})$ denote the conjugate, transpose, Hermitian (conjugate transpose), pseudo-inverse, Frobenius norm, trace, and vectorization of matrix $\mathbf{X}$, respectively. $||\mathbf{x}||_{2}$ denotes the $l_{2}$-norm of vector $\mathbf{x}$. $\mathfrak{R}\{x\}$ denotes the real part of a complex value. $\mathbb{E}\{\cdot\}$ denotes the expectation of a random variable. $|\cdot|$ and $\angle\left(\cdot\right)$ denote the modulus and angle of a complex number, respectively. $[\mathbf{x}]_{m}$ denotes the $m$-th element of vector $\mathbf{x}$, and $[\mathbf{X}]_{m,n}$ denotes the $(m,n)$-th element of matrix $\mathbf{X}$. The $m$-th row  of  matrix $\mathbf{X}$ is denoted by $[\mathbf{X}]_{m,:}$.  $\odot$ denotes the Hadamard product. $\mathrm{Diag}(\mathbf{x})$  is a  diagonal matrix having  the entries of vector $\mathbf{x}$ on its main diagonal.  Additionally, $\mathbb{C}$ and $\mathbb{R}$ refer to the fields of complex and real numbers, respectively, and $j\triangleq\sqrt{-1}$ is the imaginary unit.

\section{System Model}

\begin{figure}
		\vspace{-1cm}
\centering \includegraphics[width=2.5in,height=1.9in]{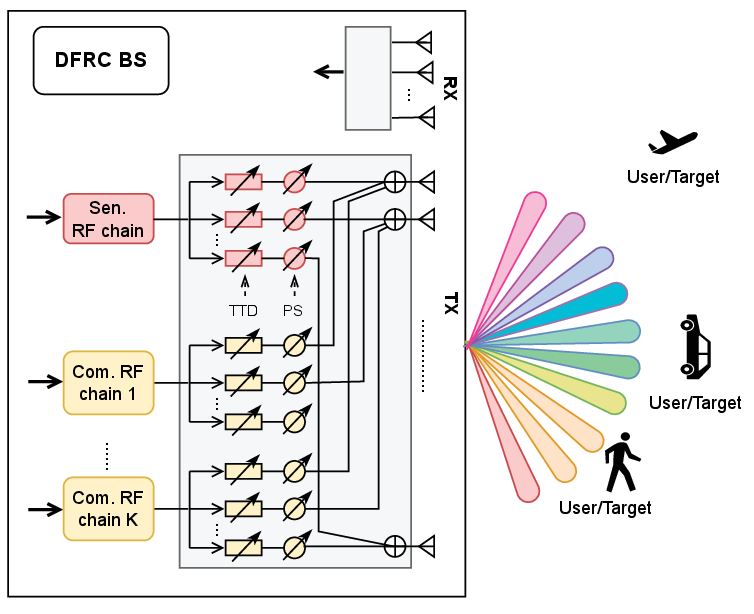}
\captionsetup{font={small}} \caption{System model of an ISAC system.}
\label{system model} 
\vspace{-1cm}
\end{figure}

In this work, we investigate an ISAC system with a specific focus on sensing-assisted communication, where  sensing is exploited for detecting and estimating the users' directions for the initial access for communication and then tracking the mobile users' positions for continously updating  the estimated angles and gains of the light-of-sight (LoSs) channels, enabling low overhead, high reliability, and low-latency communications.

Specifically, we consider a wideband mmWave massive multiple-input multiple-output (MIMO) system serving $K$ mobile communication users equipped with single antenna. Let $\theta_{k}$, $v_{k}$, and $D_{k}$ denote the angle, velocity, and distance of the $k$-th user relative to the BS, respectively. The DFRC  BS  is equipped with an  $N_{r}$-antenna uniform linear array (ULA) for reception and an  $N_{t}$-antenna ULA for transmission, where the two separate antenna arrays are used to suppress the self-interference caused by simultaneous transmission and reception.   As shown in Fig. \ref{system model},  the DFRC BS transmit antenna array is connected to $K$ RF chains providing communication services for $K$ users and an additional RF chain dedicated to sensing.  Each RF chain is connected to all $N_{t}$ transmit antennas via  $N_{t}$ PSs and $N_{t}$  TTDs.  

\subsection{Signal Model}

The considered cyclic prefix (CP)-OFDM system utilizes a total bandwidth of  $B_{{\rm tot}}=B_{{\rm sen}}+B_{\mathrm{guard}}+B_{{\rm com}}$, which is partitioned into $B_{{\rm sen}}$ for sensing, $B_{{\rm com}}$ for  communication, and $B_{\mathrm{guard}}$ for the frequency guard interval between the sensing and communication frequency bands. This frequency division  can effectively avoid mutual interference between concurrently transmitted sensing and communication signals.  In addition, bandwidth $B_{{\rm sen}}+B_{\mathrm{guard}}$ can be used to also transmit communication data when sensing is not performed and only  communication is activated. We note that different OFDM symbols are modulated separately for sensing and communication.

\subsubsection{Signal Model for Sensing}

The subcarrier spacing of the sensing OFDM signal is $\triangle f_{{\rm sen}}$,
resulting in the elementary symbol duration, $T_{{\rm sen}}=1/\triangle f_{{\rm sen}}$.
The number of subcarriers is given by $M=B_{{\rm sen}}/\triangle f_{{\rm sen}}+1$ and
assumed to be an odd number. The cyclic prefix length satisfies $N_{\mathrm{cp}}=\varrho M$,
where $\varrho=1/4\textrm{ or }1/8$ in  5G NR \cite{5gNR-2}. Define set $\mathcal{M}=\{0,...,M-1\}$. The baseband OFDM sensing signal is given by $\sum_{m \in \mathcal{M}}s_{m}(t)\triangleq  \sum_{m \in \mathcal{M}}S_{m}e^{j2\pi f_{m}t}$ \cite[Ch.  4.2.1.2]{MIMO-OFDM}, 
where $f_{m}=-\frac{B_{{\rm sen}}}{2}+m\triangle f_{{\rm sen}}$ is
the $m$-th baseband subcarrier frequency and $S_{m}$ is the complex   phase-shift keying (PSK) modulated pilot symbol on the $m$-th subcarrier.    After being up-coverted with carrier frequency
$f_{\mathrm{c}}$, the RF signal for transmission is given by $\sum_{m \in \mathcal{M}}\mathfrak{R}\left\{ \tilde{s}_{m}(t)\right\} 
\triangleq  \sum_{m \in \mathcal{M}}\mathfrak{R}\left\{ e^{j2\pi f_{\mathrm{c}}t}s_{m}(t)\right\}    \triangleq 
\sum_{m \in \mathcal{M}}\mathfrak{R}\left\{ S_{m}e^{j2\pi(f_{\mathrm{c}}+f_{m})t}\right\}$ \cite[Ch. 4.2.1.2]{MIMO-OFDM}. Denoting
the transmit (TX) analog precoding vector for the $m$-th subcarrier
by $\mathbf{f}_{m}=\mathbf{f}_{\mathrm{PS}}\odot\mathbf{t}_{m}\in\mathbb{C}^{N_{t}\times1}$
comprising a frequency-independent phase shift vector $\mathbf{f}_{\mathrm{PS}}$
and a frequency-dependent TTD vector $\mathbf{t}_{m}$, the transmitted
sensing signal can be expressed as 
\begin{align}
\sqrt{P_{{\rm s}}}\sum_{m \in \mathcal{M}}\mathbf{f}_{m}\mathfrak{R}\left\{ S_{m}e^{j2\pi(f_{\mathrm{c}}+f_{m})t}\right\},
\end{align}
where $P_{{\rm s}}$ is the transmit power.

\subsubsection{Signal Model for Communication}

To avoid multi-user interference,  we allocate different OFDM subcarriers to different users. In particular, a resource block comprising $Q$ subcarriers and bandwidth $B_{{\rm com}}/K$   is allocated to each user.  
Specifically,  the  baseband frequency of the $q$-th subcarrier allocated
to the $k$-th user is given by $f_{k,q}^{\rm com}=\frac{B_{{\rm sen}}}{2}+B_{\mathrm{guard}}+(Q(k-1)+q)\triangle f_{\mathrm{com}}$, $1\leq k\leq K$, $1\leq q\leq Q$, where $\triangle f_{\mathrm{com}}$
is the subcarrier spacing for the communication OFDM symbol\footnote{In 5G NR\cite{5gNR}, the subcarrier spacing for  data transmission 
and control signaling may be different. }   whose duration of an OFDM symbol  including the CP is defined as  $T_{\rm sym}^{\rm com}$. It is assumed that the channel between the BS and the $k$-th user
follows Rician fading comprising a  LoS component
and $L_{k}$ NLoS propagation paths. The space-frequency channel
response of the $q$-th subcarrier for the $k$-th user on the $j$ OFDM symbol is giving by  
$\mathbf{h}_{k,q,j}  =\sqrt{\frac{\rho_{k}}{\rho_{k}+1}}\alpha_{k,q,j}\mathbf{a}_{N_{t}}\left(\theta_{k},f_{{\rm c}}+f_{k,q}^{\rm com}\right)+\sqrt{\frac{1}{\rho_{k}+1}}\sum_{l=1}^{L_{k}}\alpha_{k,q,l,j}\mathbf{a}_{N_{t}}\left(\theta_{k,l},f_{{\rm c}}+f_{k,q}^{\rm com}\right)$,   with $\rho_{k}$ being the Rice factor, $\mathbf{a}_{N_{t}}$ is the steering vector, and $\theta_{k}$ and $\theta_{k,l}$ denoting the angles of departure (AoDs) of the LoS and the $l$-th NLoS path, respectively. Furthermore, $\alpha_{k,q,j}$ and $\alpha_{k,q,l,j}$ represent the frequency-dependent path fading coefficients. For example, for the LoS path, we have  $\alpha_{k,q,j}=\alpha_{k}e^{j2\pi{f_{k}^{\mathrm{d,com}}jT_{\rm sym}^{\rm com}}}e^{-j2\pi f_{k,q}^{\rm com}\tau_{k}^{\mathrm{com}}}$ with frequency-independent fading coefficient $\alpha_{k}$,  Doppler frequency $f_{k}^{\mathrm{d,com}}=f_{\mathrm{c}}\frac{v_{k}}{c}$, and delay $\tau_{k}^{\mathrm{com}}=\frac{D_{k}}{c}$ for one-way communication.  Since the main focus of this paper is the estimation of the angles, velocities, and distances via rainbow beams, for evaluation of the resulting communication performance in Section \ref{simulation}, we assume pure analog precoding.  The  analog precoding vector is given by  $\mathbf{w}_{k,q}={\bf w}_{\mathrm{PS},k}\odot\mathbf{t}_{k,q}^{{\rm com}}$
containing a frequency-independent phase shift vector, ${\bf w}_{\mathrm{PS},k}$,
and a frequency-dependent TTD vector, $\mathbf{t}_{k,q}^{{\rm com}}$,
specific to the $q$-th subcarrier. Then, the achievable data rate of the $k$-th
user on the $q$-th subcarrier is given by 
\begin{align}
R_{k,q}=\log_{2}\left(1+\frac{P_{{\rm c},k}}{\sigma_{{\rm c},k}^{2}}|\mathbf{h}_{k,q}^{\mathrm{H}}{\bf w}_{k,q}|^{2}\right),\label{eq:hi86}
\end{align}
where $P_{{\rm c},k}$ and $\sigma_{{\rm c},k}^{2}$ are the transmit and noise powers for user $k$ per subcarrier, respectively. 

Optimizing   $\mathbf{w}_{k,q}$ for rate maximization  requires the channel state information (CSI) of $\mathbf{h}_{k,q}$, which is typically obtained via pilot-based channel estimation methods.  However, given the rapid time-varying nature of the channels in mobile scenarios and the dominance of the LoS path in high-frequency systems, it is desirable to leverage radar sensing to acquire CSI for the LoS path comprising angle, velocity, and distance.  Based on the estimated CSI of the LoS path, $\mathbf{w}_{k,q}$ can be designed by neglecting the NLoS components.

\subsection{Controllable Beam Squint}

Prior to delving into the design of precoding vectors, $\mathbf{f}_{m}$ and $\mathbf{w}_{k,q}$, it is imperative to highlight the frequency-dependent properties of the massive antenna array for wideband mmWave
signals, i.e., the beam squint effect. Let us consider the  sensing OFDM signal as an example. Consider an angle $\theta \in [0,\pi]$ measured
with respect to (w.r.t.) the array plane. The frequency-dependent steering vector in the far field of the array at angle $\theta$ and frequency
$f_{m}+f_{\mathrm{c}}$ is given by $\mathbf{a}_{N_{t}}(\theta,f_{\mathrm{c}}+f_{m})  =[1,e^{-j2\pi(f_{\mathrm{c}}+f_{m})\frac{d}{c}\cos\theta},\ldots,e^{-j2\pi(N_{t}-1)(f_{\mathrm{c}}+f_{m})\frac{d}{c}\cos\theta}]^{\mathrm{T}}$,
where $d$ and $c$ represent the antenna spacing and the speed of light,  respectively. When the propagation delay across the array aperture
is less than the sampling duration, i.e., $(N_{t}-1)\frac{d}{c}<\frac{1}{B_{{\rm sen}}}$,
$\mathbf{a}_{N_{t}}(\theta,f_{m}+f_{\mathrm{c}})$ can be approximated
as a narrowband steering vector, i.e., $\mathbf{a}_{N_{t}}(\theta,f_{\mathrm{c}})$
\cite{2019Asilomar}\footnote{In LTE/5G NR, the mmWave frequency band covers 30-300 GHz with bandwidths of 400 MHz or 800 MHz \cite{LTE,5gNR}.  A system with $f_{\mathrm{c}}=30$
GHz, $B_{{\rm sen}}=400$ MHz, and $d=\frac{c}{2f_{\mathrm{c}}}$ can
be regarded as a narrowband system when $N_{t}<\frac{2f_{\mathrm{c}}}{B_{{\rm sen}}}+1=151$.}. In contrast,  for a given carrier frequency $f_{\mathrm{c}}$, a large number of antennas $N_{t}$ or a large bandwidth $B_{{\rm sen}}$  lead to the beam squint effect. The beam squint effect causes the beams attributed to different subcarriers to be directed towards distinct  angles if a frequency-independent precoding vector is employed. 

Specifically, define the beampattern of phase shift precoder $\mathbf{f}_{\mathrm{PS}}$
at angle $\theta$ and frequency $f_{m}+f_{\mathrm{c}}$ as $\left|\mathbf{a}_{N_{t}}^{\mathrm{H}}\left(\theta,f_{m}+f_{\mathrm{c}}\right)\mathbf{f}_{\mathrm{PS}}\right|^{2}$.
If we let the beam of the center subcarrier, with frequency $f_{\mathrm{c}}$,
point towards direction angle $\theta_{\mathrm{c}}$, precoder $\mathbf{f}_{\mathrm{PS}}$
should be chosen as $\mathbf{f}_{\mathrm{PS}}=\mathbf{a}_{N_{t}}\left(\theta_{\mathrm{c}},f_{\mathrm{c}}\right)$,
such that $\left|\mathbf{a}_{N_{t}}^{\mathrm{H}}\left(\theta,f_{\mathrm{c}}\right)\mathbf{f}_{\mathrm{PS}}\right|^{2}$
has the maximum value $N_{t}^{2}$ when $\theta=\theta_{\mathrm{c}}$.
Given the  frequency-independent precoder, the main lobe peak of the
beampattern 
\begin{align}
\left|\mathbf{a}_{N_{t}}^{\mathrm{H}}\left(\theta,f_{m}+f_{\mathrm{c}}\right)\mathbf{f}_{\mathrm{PS}}\right|^{2} & =\left|\sum_{n=1}^{N_{t}}e^{j2\pi(n-1)\left((f_{\mathrm{c}}+f_{m})\frac{d}{c}\cos\theta-f_{\mathrm{c}}\frac{d}{c}\cos\theta_{\mathrm{c}}\right)}\right|^{2}
\end{align}
for the subcarrier at frequency $f_{m}+f_{\mathrm{c}}$ is located
at 
\begin{align}
\theta_{m}^{\mathrm{PS}} & =\arccos\left(\frac{f_{\mathrm{c}}}{f_{\mathrm{c}}+f_{m}}\cos\theta_{\mathrm{c}}\right)\label{eq:ki9}
\end{align}
satisfying $(f_{\mathrm{c}}+f_{m})\frac{d}{c}\cos\theta_{m}^{\mathrm{PS}}-f_{\mathrm{c}}\frac{d}{c}\cos\theta_{\mathrm{c}}=0$.
In narrowband systems with $\frac{f_{m}}{f_{\mathrm{c}}}\ll1$, a
highly focused OFDM beam with one main lobe peak direction at angle
$\theta_{\mathrm{c}}$ can be generated to serve the users. However,
in wideband systems, the main lobe peak $\theta_{m}^{\mathrm{PS}}$,
for non-center carriers, deviates from the desired direction $\theta_{\mathrm{c}}$.
This may lead to performance degradation. To control the beam squint
effect in wideband communication systems, TTDs are utilized, which introduce constant delays to all subcarriers, resulting
in a frequency-dependent steering vector that enables control of the
beam direction for different signal frequencies. In this work, we consider the commonly used RF TTDs and denote by  $\kappa_{n}=(n-1)\triangle\kappa$
 the delay of the delay line for  the $n$-th transmit antenna, where
$\triangle\kappa$ is the delay spacing between adjacent antennas.
This  produces frequency-dependent TTD vectors, as $\mathbf{t}_{m}=[1,e^{-j2\pi f_{m}\triangle\kappa},\ldots,e^{-j2\pi f_{m}(N_{t}-1)\triangle\kappa}]^{\mathrm{T}}$.
Then, the beampattern of $\mathbf{f}_{m}=\mathbf{f}_{\mathrm{PS}}\odot\mathbf{t}_{m}$
is given by 
\begin{align}
G(\theta,f_{m}) & =\left|\mathbf{a}_{N_{t}}^{\mathrm{H}}\left(\theta,f_{m}+f_{\mathrm{c}}\right)(\mathbf{f}_{\mathrm{PS}}\odot\mathbf{t}_{m})\right|^{2}\nonumber \\
 & =\left|\sum_{n=1}^{N_{t}}e^{j2\pi(n-1)\left((f_{\mathrm{c}}+f_{m})\frac{d}{c}\cos\theta-f_{\mathrm{c}}\frac{d}{c}\cos\theta_{\mathrm{c}}-f_{m}\triangle\kappa\right)}\right|^{2}\label{eq:koe}
\end{align}
with the main lobe peak of subcarrier $m$ pointing to 
\begin{align}
\theta_{m}^{\mathrm{TTD}} & =\arccos\left(\frac{\frac{c}{d}f_{m}\triangle\kappa+f_{\mathrm{c}}\cos\theta_{\mathrm{c}}}{f_{\mathrm{c}}+f_{m}}\right).\label{eq:se9}
\end{align}

\subsubsection{Mitigation of Beam Squint for Communication}
\label{beamscom}

For communication, the beam squint effect must be mitigated such that the beams of  all subcarriers focus towards the user of interest.  This can be achieved by designing the delay parameter as $\triangle\kappa=\frac{d}{c}\cos\theta_{\mathrm{c}}$ such that $\theta_{m}^{\mathrm{TTD}}$ in (\ref{eq:se9}) is independent of $f_{m}$. 

\subsubsection{Utilization of Beam Squint for Sensing}

For sensing, the beam squint effect can be strategically harnessed,  allowing the beams corresponding to different subcarriers to be concurrently directed toward different directions by properly choosing the TTD vector. The resulting set of OFDM subcarrier beams are called ``rainbow beams", encompassing all potential positions of the sensing targets simultaneously, as elaborated in Section \ref{section iii}.

\subsection{Frame Structure}

\begin{figure}[htbp]
	\vspace{-0.5cm}
	\begin{minipage}[t]{0.65\textwidth}%
		\centering \includegraphics[width=4.2in,height=1.2in]{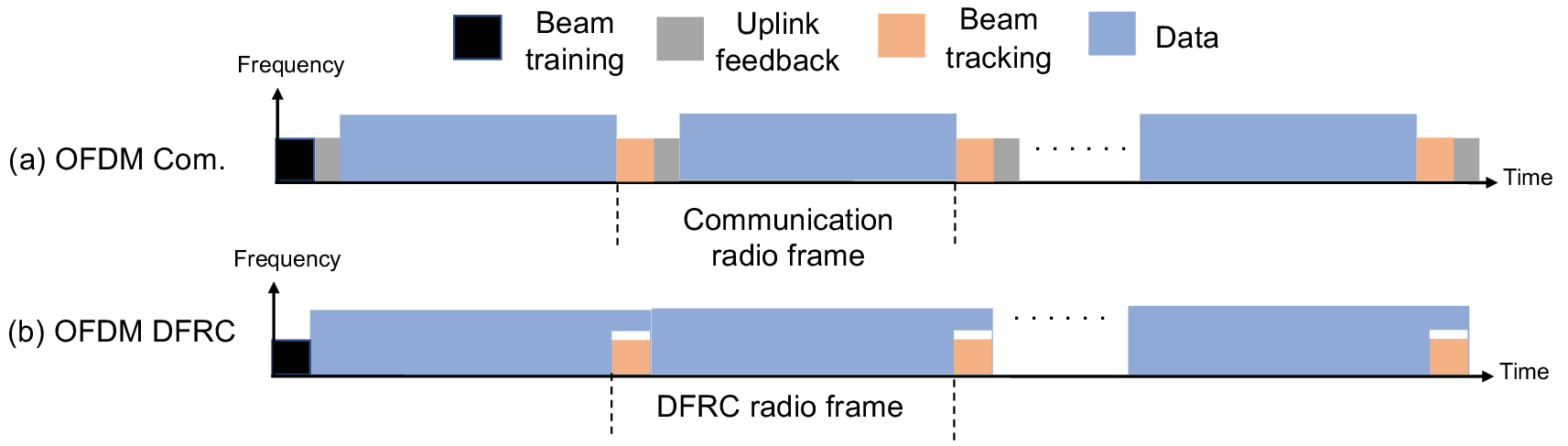}
		\captionsetup{font={small}} \caption{Frame structure of the proposed rainbow beam aided ISAC.}
		\label{frame} 
	\end{minipage}\centering %
	\begin{minipage}[t]{0.35\textwidth}%
		\centering \includegraphics[width=2.2in,height=1.2in]{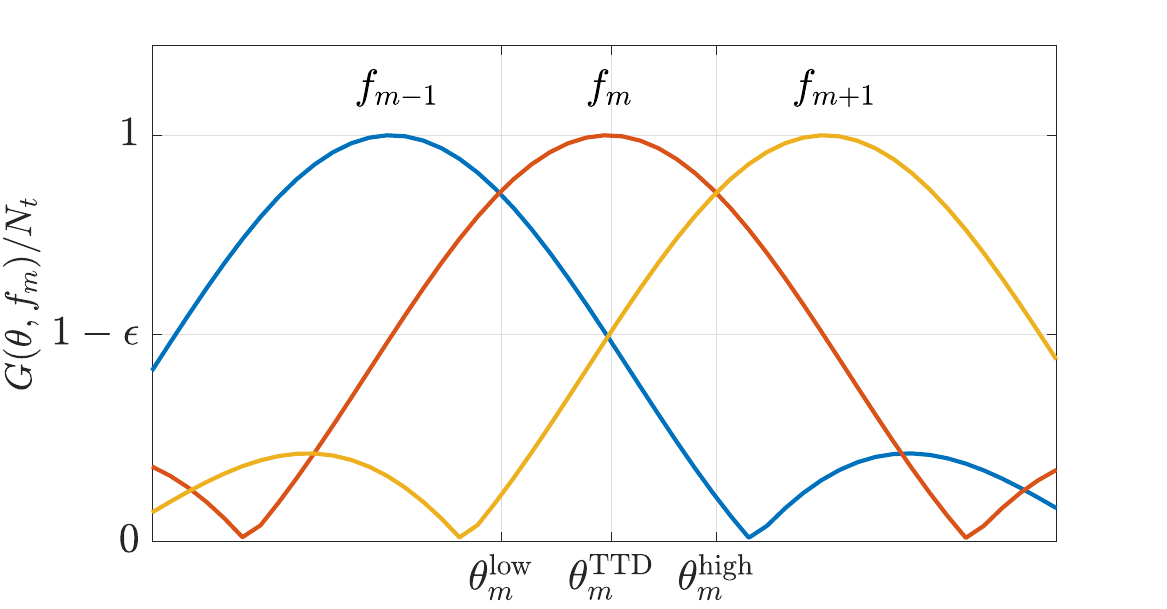}
		\captionsetup{font={small}}
		\caption{ Example of overlapping beams. }
		\label{5db-overlapped beam} %
	\end{minipage}
\vspace{-1cm}
\end{figure}

To establish reliable communication links, accurate information regarding the directions of the users at the BS is essential. Conventionally,
beam training is employed by the BS to identify the users' directions for initial access, achieved through scanning the entire angular space using narrow beams. Subsequently, to maintain the connection, the users' directions have to be tracked by scanning a smaller angular space that covers the previously
identified directions \cite{TrainingTracking}. However, this approach  inevitably introduces significant latency and communication overhead,
as it necessitates substantial pilot and feedback overhead.   The corresponding  frame structure is shown  in Fig. \ref{frame}(a). Furthermore, this approach  requires frequent and fast switching of the  array PSs, which is difficult to realize at high frequencies \cite{rainbow-TCS2021}. To reduce the overhead and hardware complexity,  and achieve high-reliability and low-latency communications, we propose a novel scheme comprising  fast beam training and continuous user tracking strategies by leveraging  the synergies between DFRC and rainbow beamforming. The proposed frame  structure for rainbow beam aided ISAC systems is depicted in Fig.  \ref{frame}(b).

\textit{1) Beam Training Block:} The DFRC BS employs  a frequency-dependent TTD array  forming rainbow beams based on an OFDM signal. These beams  are designed to  cover the entire angular space,  enabling the detection and estimation of angles, distances, and velocities of communication users.

\textit{2) Simultaneous Beam Tracking And Communication Block: }The DFRC BS effectively combines the functionalities of downlink  communication and movement trajectory sensing. This integrated operation enables
the system to perform seamless tracking of users while continuously  communicating with them simultaneously in a different frequency band\cite{LiuFan-TWC2020}. For communication, the BS designs the communication beamforming based on the previously estimated user  angles. Notably, the rainbow beam enabled training conducted during the initial access block is highly efficient,  requiring minimal OFDM  resources and no hardware overhead for  PS and TTD switching.  We  assume that a user's direction information obtained in the initial access block remains unchanged during the subsequent radio frames\footnote{For instance, assuming a user with velocity 50 m/s is located 350 m  from the BS, then the azimuth angle change is 0.0014° during a  radio frame of 10 ms. Thus,  the azimuth angle can be reasonably approximated as remaining constant for two radio frames.}. 
Furthermore, we approximate the user's estimated position information derived from radar signal reflection to be equal to its antenna position, valid in far-field communication scenarios where a user with small cross-section area is situated at a considerable distance from the BS and can be treated as a point target \cite{LiuFan-TSP}.
Regarding user tracking,  the DFRC-BS employs one RF chain to create multiple rainbow beam sectors, each comprising  adjacent subcarrier beams and covering a small angular space,  for simultaneous monitoring of multiple users.
Users having similar angular directions are tracked within the same sector. Due to the shared time resources between user tracking and communication, the proposed scheme enables uninterrupted communication, leading to enhanced  spectral effeciency  and reduced system-level communication latency. 

\section{Design of Rainbow Beams for Sensing}

\label{section iii}

In this subsection, the beam squint effect is exploited to realize fast sensing of multiple users. To this end, two essential problems must
be tackled for the design of rainbow beams. Firstly, the delays of the TTDs have to be designed to ensure  full angular coverage of  the beams corresponding to different subcarriers.  Secondly, the subcarrier spacing and total number of sensing subcarriers should be determined to guarantee angle resolution, ensuring that any arbitrary angle is  covered by at least one subcarrier with sufficient beam gain for robust detection.

\subsection {Problem 1: Coverage}

Let  $[\theta_{\mathrm{low}},\theta_{\mathrm{high}}]\subseteq[0,\pi]$ define  the angular space covering all potential user directions. The aim of the TTD delay design is to let the beams of all subcarriers  $f_{m\in \mathcal{M}}\in[-\frac{B_{{\rm sen}}}{2},\frac{B_{{\rm sen}}}{2}]$ cover the entire angular space. To this end, we let subcarriers $f_{0}=-\frac{B_{{\rm sen}}}{2}$ and $f_{M-1}=\frac{B_{{\rm sen}}}{2}$ points to $\theta_{\mathrm{low}}$ and $\theta_{\mathrm{high}}$, respectively, and obtain the  following lemma.
\begin{lemma}\label{lemmxx} To sense users within the angular space $[\theta_{\mathrm{low}},\theta_{\mathrm{high}}]\subseteq[0,\pi]$
using a multi-carrier signal with bandwidth $B_{{\rm sen}}$ and carrier frequency
$f_{{\rm c}}$, the PSs and TTDs can be designed by determining $\triangle\kappa$
and $\theta_{\mathrm{c}}$ as follows: 
\begin{align}
\triangle\kappa & =\frac{1}{2}\left[\left(\frac{1}{B_{{\rm sen}}}+\frac{1}{2f_{\mathrm{c}}}\right)\cos\theta_{\mathrm{high}}-\left(\frac{1}{B_{{\rm sen}}}-\frac{1}{2f_{\mathrm{c}}}\right)\cos\theta_{\mathrm{low}}\right],\label{eq:f45-1}\\
\theta_{\mathrm{c}} & =\arccos\left(\frac{B_{{\rm sen}}}{2}\left[\left(\frac{1}{B_{{\rm sen}}}+\frac{1}{2f_{\mathrm{c}}}\right)\cos\theta_{\mathrm{high}}+\left(\frac{1}{B_{{\rm sen}}}-\frac{1}{2f_{\mathrm{c}}}\right)\cos\theta_{\mathrm{low}}\right]\right).\label{eq:1e5-1}
\end{align}
\end{lemma} 
\textbf{\textit{Proof: }}  Lemma \ref{lemmxx} holds under the assumption of half-wavelength
antenna spacing, i.e., $\frac{c}{d}=2f_{\mathrm{c}}$, and can be
proved by combining $\cos\theta_{\rm low}=\frac{2f_{\mathrm{c}}f_{0}\triangle\kappa+f_{\mathrm{c}}\cos\theta_{\mathrm{c}}}{(f_{\mathrm{c}}+f_{0})}$
and $\cos\theta_{\rm high}=\frac{2f_{\mathrm{c}}f_{M+1}\triangle\kappa+f_{\mathrm{c}}\cos\theta_{\mathrm{c}}}{(f_{\mathrm{c}}+f_{M+1})}$. \hspace{0.01cm}$\blacksquare$

\subsection{Problem 2: Resolution}

The resolution of the rainbow beams depends on the total number of subcarriers used. A larger number of subcarriers provides a higher angular
resolution, which however causes challenges for detection and estimation
due to the associated high sampling rates and main lobe ambiguities.
Herein, our initial focus is on ascertaining the minimal required
number of subcarriers. To avoid missing users due to uncovered directions, following\cite{2019Asilomar}, overlapping beams are needed, such that
for any arbitrary direction, at least one subcarrier has adequate beamforming
gain. However, compared to  [14], the scope of this work is expanded to the estimation of  additional parameters such as the users' distances and velocities. Therefore, the design goal for the
overlapping beams is to ensure that, for any arbitrary angle,  at least two beams provide a power gain of no less than $(1-\epsilon)N_{t}^{2}$,
where $\epsilon\in[0,1]$ is the reduction in power compared to the
maximum  power, i.e., $N_{t}^{2}$, and is referred to as beam overlap factor in this paper. In other words, any arbitrary angle should
be covered by the $\epsilon$-beamwidth of at least two beams. Mathematically, define the angular space $[\theta_{m}^{\mathrm{low}},\theta_{m}^{\mathrm{high}}]$, 
wherein the beam of $f_{m}$ provides a larger gain compared with the  adjacent beams, as shown in Fig. \ref{5db-overlapped beam}. The design problem
is to identify the feasible set $\mathcal{S}$ of design parameters,
ensuring that within region $[\theta_{m}^{\mathrm{low}},\theta_{m}^{\mathrm{high}}]$,
the beam gain of at least one adjacent subcarrier does not drop
below $(1-\epsilon)N_{t}^{2}$: 
\begin{equation}
\mathcal{S}=\left\{ (\triangle f_{{\rm sen}},M)\left|\left(\min_{1\leq m\leq M}\min_{\theta\in[\theta_{m}^{\mathrm{low}},\theta_{m}^{\mathrm{high}}]}\max(G(\theta,f_{m+1}),G(\theta,f_{m-1}))\right)\geq(1-\epsilon)N_{t}^{2}\right.\right\} .\label{eq:j9i8}
\end{equation}

As shown in Fig. \ref{5db-overlapped beam}, for a given $f_{m}$,
the main lobes of adjacent subcarriers $f_{m-1}$ and $f_{m+1}$ intersect
at the angle of $\theta_{m}^{\mathrm{TTD}}\in[\theta_{m}^{\mathrm{low}},\theta_{m}^{\mathrm{high}}]$,
yielding $G(\theta_{m}^{\mathrm{TTD}},f_{m+1})=G(\theta_{m}^{\mathrm{TTD}},f_{m-1})$.
This implies that in the angular space $[\theta_{m}^{\mathrm{low}},\theta_{m}^{\mathrm{high}}]$,
the minimum gain for the adjacent subcarriers is obtained for  $\theta_{m}^{\mathrm{TTD}}$.
Then, Problem (\ref{eq:j9i8}) becomes 
\begin{equation}
\mathcal{S}=\left\{ (\triangle f_{{\rm sen}},M)\left|\left(\min_{1\leq m\leq M}G(\theta_{m}^{\mathrm{TTD}},f_{m+1})\right)\geq(1-\epsilon)N_{t}^{2}\right.\right\} .\label{eq:j9i80l}
\end{equation}

By utilizing (\ref{eq:koe}) and defining 
\begin{align}
\psi(\theta_{m}^{\mathrm{TTD}},f_{m+1}) & =(f_{\mathrm{c}}+f_{m+1})\frac{d}{c}\cos\theta_{m}^{\mathrm{TTD}}-f_{\mathrm{c}}\frac{d}{c}\cos\theta_{\mathrm{c}}-f_{m+1}\triangle\kappa\nonumber \\
 & =\frac{\cos\theta_{\mathrm{c}}-2f_{\mathrm{c}}\triangle\kappa}{2(f_{\mathrm{c}}+f_{m})}\triangle f_{{\rm sen}},
\end{align}
(\ref{eq:j9i80l}) is equivalent to 
\begin{equation}
\mathcal{S}=\left\{ (\triangle f_{{\rm sen}},M)\left|\left(\max_{1\leq m\leq M}2\pi\left|\psi(\theta_{m}^{\mathrm{TTD}},f_{m+1})\right|\right)\leq\frac{\triangle\psi_{\epsilon}}{2}\right.\right\} ,\label{eq:j9i86}
\end{equation}
where $\triangle\psi_{\epsilon}$ is the $\epsilon$-beamwidth and
can be found numerically \cite[Ch. 22.7]{EW-antenna}. For example,
the 3dB-beamwidth is $\triangle\psi_{\mathrm{3dB}}=0.886(2\pi/N_{t})$.
The solution of Problem (\ref{eq:j9i86}) is provided  in the following
lemma. \begin{lemma}\label{lemmyy} For rainbow beams
with  beam overlap factor ${\epsilon}$ to cover an angular space
of $[\theta_{\mathrm{low}},\theta_{\mathrm{high}}]\subseteq[0,\pi]$
using a multi-carrier signal with bandwidth $B_{{\rm sen}}$ and carrier
frequency $f_{{\rm c}}$, the subcarrier spacing and the number
of subcarriers need to satisfy 
\begin{align}
\triangle f_{{\rm sen}} & \leq\frac{\triangle\psi_{\epsilon}}{4\pi\left(\cos\theta_{\mathrm{c}}\left/(2f_{\mathrm{c}})\right.-\triangle\kappa\right)} =\frac{\triangle\psi_{\epsilon}}{\frac{2\pi}{B_{{\rm sen}}}(1-\frac{B_{{\rm sen}}^{2}}{4f_{\mathrm{c}}^{2}})(\cos\theta_{\mathrm{low}}-\cos\theta_{\mathrm{high}})},\label{eq:e3j9y}\\
M & =\frac{B_{{\rm sen}}}{\triangle f_{{\rm sen}}}+1 \geq\frac{2\pi(1-\frac{B_{{\rm sen}}^{2}}{4f_{\mathrm{c}}^{2}})(\cos\theta_{\mathrm{low}}-\cos\theta_{\mathrm{high}})}{\triangle\psi_{\epsilon}}+1.\label{eq:j987}
\end{align}
\end{lemma} \textbf{\textit{Proof: }} It is obvious that the maximum
value of $\left|\psi(\theta_{m}^{\mathrm{TTD}},f_{m+1})\right|$ is
obtained when $f_{m}=0$, leading to 
\begin{align}
\frac{\triangle\psi_{\epsilon}}{2} & \geq2\pi\left|\left(\frac{\cos\theta_{\mathrm{c}}}{2f_{\mathrm{c}}}-\triangle\kappa\right)\triangle f_{{\rm sen}}\right|.\label{eq:y7u8}
\end{align}
Combining (\ref{eq:f45-1}) and (\ref{eq:1e5-1}), we have $\frac{\cos\theta_{\mathrm{c}}}{2f_{\mathrm{c}}}-\triangle\kappa=\frac{B_{{\rm sen}}}{2}(\frac{1}{B_{{\rm sen}}^{2}}-\frac{1}{4f_{\mathrm{c}}^{2}})(\cos\theta_{\mathrm{low}}-\cos\theta_{\mathrm{high}})>0$.
Therefore, the maximum subcarrier spacing  given in (\ref{eq:e3j9y})
can be derived directly from (\ref{eq:y7u8}). \hspace{2.5cm}$\blacksquare$

For example, for a system with $\frac{B_{{\rm sen}}^{2}}{4f_{\mathrm{c}}^{2}}\ll1$
and an angular space of $[\theta_{\mathrm{low}},\theta_{\mathrm{high}}]=[0,\pi]$,
the critical values for the subcarrier spacing and the  required number of subcarriers
are $\triangle f_{{\rm sen}}=\frac{\triangle\psi_{\epsilon}B_{{\rm sen}}}{4\pi}$
and $M=\left\lceil 4\pi/\triangle\psi_{\epsilon}\right\rceil +1$,
respectively, where ceiling operator $\left\lceil x\right\rceil $
gives the smallest integer that is greater than $x$. 

\section{Rainbow Beam Training And Tracking}

In this section, we first develop a model for the received signal for rainbow beam training, before proposing three parameter estimation schemes. Finally,  based on the constructed framework, rainbow beam-based user tracking is established avoiding switching of  PSs and TTDs, which are typically necessary in existing tracking technologies.

\subsection{Beam Training - Received Signal Model }

In this section, we model the echo signal of the rainbow beams emitted
by the DFRC BS.  Let $m_{k}$ denote the index of the subcarrier reflected by the
$k$-th user. To avoid missing any user within the angular space as elaborated in Section \ref{section iii},
the sensing rainbow beams must overlap, which causes a given user  to be located in the main lobes of the beams created by multiple  adjacent subcarriers. We consider a small $N_{r}$ from a cost-effective standpoint, resulting in a frequency-independent  receive steering vector $\mathbf{a}_{N_{r}}(\theta_{k})=[1,e^{-j2\pi f_{\mathrm{c}}\frac{d}{c}\cos\theta_{k}},\ldots,e^{-j2\pi(N_{r}-1)f_{\mathrm{c}}\frac{d}{c}\cos\theta_{k}}]^{\mathrm{T}}$. The angle of arrival (AoA), $\theta_{k}$ is identical to the AoD in the transmit steering vector, $\mathbf{a}_{N_{t}}(\theta_{k},f_{m}+f_{\mathrm{c}})$.   Define    set $\mathcal{M}_{k}$ comprising the indices of the subcarriers whose beams are reflected by the $k$-th user with a relatively high gain, providing useful information for estimation\footnote{$\mathcal{M}_{k}$ can be determined by $\mathcal{M}_{k}=\left\{ \left.m\right||\mathbf{a}_{N_{t}}^{\mathrm{H}}\left(\theta_{k},f_{\mathrm{c}}+f_{m}\right)\mathbf{f}_{m}|^{2}>\epsilon_{\mathcal{M}}, m\in \mathcal{M}\right\}$, where $\epsilon_{\mathcal{M}}$ is a threshold. }.  
 Collecting the passband echo signal from the $K$ users,  the corresponding down-converted complex  baseband   signal is given by 
\begin{align}
{\bf y}(t) 
  =\sum_{k=1}^{K}\mathbf{a}_{N_{r}}\left(\theta_{k}\right)\left(\sum_{m_{k}\in\mathcal{M}_{k}}z_{m_{k}}(t)+\sum_{m_{k}\in\mathcal{M}\setminus\mathcal{M}_{k}} z_{m_{k}}(t)\right)+\mathbf{n}_{r}(t),\label{eq:ji45}
\end{align}
where   
\begin{align}z_{m_{k}}(t)&=e^{-j2\pi f_{\mathrm{c}}t}\sqrt{P_{{\rm s}}}\beta_{k}\mathbf{a}_{N_{t}}^{\mathrm{H}}\left(\theta_{k},f_{\mathrm{c}}+f_{m_{k}}\right)\mathbf{f}_{m_{k}}\widetilde{s}_{m_{k}}\left(t-\frac{2(D_{k}-v_{k}t)}{c}\right)\nonumber\\
&=\sqrt{P_{{\rm s}}}\beta_{k}\mathbf{a}_{N_{t}}^{\mathrm{H}}\left(\theta_{k},f_{\mathrm{c}}+f_{m_{k}}\right)\mathbf{f}_{m_{k}}e^{j2\pi f_{k}^{\mathrm{d}}t}e^{-j2\pi f_{\mathrm{c}}\tau_{k}}S_{m}e^{j2\pi f_{m_k}t}s_{m_{k}}\left(\left(1+\frac{2v_{k}}{c}\right)t-\tau_{k}\right).\label{eq:zz54}
\end{align}
Here, $\sum_{m_{k}\in\mathcal{M}_{k}}z_{m_{k}}(t)$ represents the useful echo signal from the  subcarriers in $\mathcal{M}_{k}$ and \\ $\sum_{m_{k}\in\mathcal{M}\setminus\mathcal{M}_{k}} z_{m_{k}}(t)$ is the interference caused by the $k$-th user on the remaining subcarriers.  Complex channel gain $\beta_{k}$ comprises the path loss and radar
cross section and has power $|\beta_{k}|^{2}=\frac{\lambda_{\mathrm{c}}^{2}\sigma_{\mathrm{rcs}}}{(4\pi)^{3}D_{k}^{4}}$
with radar cross section $\sigma_{\mathrm{rcs}}$ and center carrier
wavelength $\lambda_{\mathrm{c}}$. Then, $f_{k}^{\mathrm{d}}=f_{\mathrm{c}}\frac{2v_{k}}{c}$ and $\tau_{k}=\frac{2D_{k}}{c}$ represent the Doppler frequency and round-trip delay of the $k$-th user, respectively. Vector $\mathbf{n}_{r}(t)$ is the  additive white Gaussian noise (AWGN) following the distribution $\mathbf{n}_{r}(t)\sim\mathcal{CN}(0,\sigma^{2}_{\rm s}\mathbf{I})$ with noise power $\sigma^{2}_{\rm s}$ at the BS.

The term $(1+\frac{2v_{k}}{c})t$ implies a time stretching/compression of  the reflection received  from the $k$-th user due to the Doppler  effect, giving rise to range migration and Doppler scaling. When the users'  maximum movement within a single OFDM symbol remains
significantly less than the distance (range) resolution, i.e., $T_{{\rm sen}}v_{k}\ll c/(2B_{{\rm sen}})$,
OFDM radar signal processing techniques typically neglect this effect,
adopting the narrowband approximation $s_{m_{k}}((1+\frac{2v_{k}}{c})t-\tau_{k})\approx s_{m_{k}}(t-\tau_{k})$
\cite{2011radarproc}. This condition is equivalent to a small time-bandwidth
product $T_{{\rm sen}}B_{{\rm sen}}\ll\frac{c}{|2v_{k}|}$ and $\left|\frac{2v_{k}}{c}\right|\ll\frac{1}{M}$
\cite{2018radarTVT}. In addition, constant $e^{-j2\pi f_{\mathrm{c}}\tau_{k}}$
can be absorbed into $\tilde{\beta}_{k}=\beta_{k}e^{-j2\pi f_{\mathrm{c}}\tau_{k}}$.
With these adjustments, (\ref{eq:zz54}) becomes 
\begin{align}
	z_{m_{k}}(t)
	&=\sqrt{P_{{\rm s}}}\tilde{\beta}_{k}\mathbf{a}_{N_{t}}^{\mathrm{H}}\left(\theta_{k},f_{\mathrm{c}}+f_{m_{k}}\right)\mathbf{f}_{m_{k}}e^{j2\pi f_{k}^{\mathrm{d}}t}s_{m_{k}}\left(t-\tau_{k}\right).\label{eq:z004}
\end{align}

After matched filtering and sampling at the sampling frequency $M\triangle f_{{\rm sen}}$ at times $t=\frac{nT_{{\rm sen}}}{M}=\frac{n}{M\triangle f_{{\rm sen}}}$, where $n=0,...,M-1$,  the discrete-time version of the down-converted signal is obtained as 
\begin{align}
{\bf y}[n]
=\sum_{k=1}^{K}\mathbf{a}_{N_{r}}\left(\theta_{k}\right)\left(\sum_{m_{k}\in\mathcal{M}_{k}}z_{m_{k}}[n]+\sum_{m_{k}\in\mathcal{M}\setminus\mathcal{M}_{k}} z_{m_{k}}[n]\right)+\mathbf{n}_{r}[n],\label{eq:jt6t65}
\end{align}
where  $z_{m_{k}}[n]
	=\sqrt{P_{{\rm s}}}\tilde{\beta}_{k}\mathbf{a}_{N_{t}}^{\mathrm{H}}\left(\theta_{k},f_{m_{k}}\right)\mathbf{f}_{m_{k}} e^{j2\pi f_{k}^{\mathrm{d}}\frac{nT_{{\rm sen}}}{M}} S_{m_{k}} e^{-j2\pi f_{m_{k}}\tau_{k}}e^{j2\pi(m_{k}-\frac{M-1}{2})\frac{n}{M}}$  
and  $\mathbf{n}_{r}[n]$ is the discrete-time version  of $\mathbf{n}_{r}(t)$.

\subsection{Beam Training - Detection and Estimation}

In this subsection, we first develop radar detection and estimation algorithms based on a single OFDM symbol and a single-antenna receiver, which can  estimate the users' parameters reliably for low Doppler frequencies but  leads to Doppler ambiguity for high Doppler frequencies. To overcome this issue, we propose two alternative schemes that exploit two OFDM symbols and a multi-antenna receiver, respectively.

\textbf{(1) Single-Antenna Receiver}
\label{SA1O}

Assuming a single-antenna receiver, (\ref{eq:jt6t65})  simplifies to ${y}[n]  =\sum_{k=1}^{K}\sum_{m_{k}\in\mathcal{M}_{k}}z_{m_{k}}[n]+i[n]+n_{r}[n]$, 
where $i[n]=\sum_{k=1}^{K}\sum_{m_{k}\in\mathcal{M}\setminus\mathcal{M}_{k}} z_{m_{k}}[n]$ and $n_{r}[n]$ is the AWGN.  
Collecting   samples $y[n] $, $n=0,...,M-1$, in measurement vector 
$\mathbf{y}=[y[0],y[1],\hdots,y[M-1]]^{{\rm T}}$, we obtain
\begin{align}
	\mathbf{y}\!=\!\mathbf{D}_{M}\!\sum_{k=1}^{K}\!\mathbf{D}(v_{k})\!\sum_{m_{k}\in\mathcal{M}_{k}}\!\!\mathbf{d}_{\mathrm{idft},m_{k}}\tilde{S}_{m_{k}}(\tau_{k})\!\!+\!\mathbf{i}\!+\!\mathbf{n},\label{eqjii0}
\end{align}
where 
\begin{subequations}
	\label{hc} 
	\begin{align}
		&\mathbf{D}_{M}  \!=\!\mathrm{Diag}(1,e^{-j\pi(M-1)\frac{1}{M}},\ldots,e^{-j\pi(M-1)\frac{(M-1)}{M}}),\label{b1}\\
		&\mathbf{D}(v_{k})  \!=\!\mathrm{Diag}(1,\!e^{j2\pi f_{k}^{\mathrm{d}}\frac{T_{{\rm sen}}}{M}},\!\ldots,\!e^{j2\pi f_{k}^{\mathrm{d}}\frac{(M-1)T_{{\rm sen}}}{M}}),\label{b2}\\
		&\tilde{S}_{m_{k}}(\tau_{k})  \!=\!\sqrt{P_{{\rm s}}}\tilde{\beta}_{k}\mathbf{a}_{N_{t}}^{\mathrm{H}}\left(\theta_{k},f_{m_{k}}\right)\mathbf{f}_{m_{k}}S_{m_{k}}e^{-j2\pi f_{m_{k}}\tau_{k}},\label{b3}\\
		&\mathbf{d}_{\mathrm{idft},m_{k}}  =[1,e^{j2\pi m_{k}\frac{1}{M}},\hdots,e^{j2\pi m_{k}\frac{(M-1)}{M}}]^{{\rm T}},\label{b4}\\
		&\mathbf{i}  \!=\![i[0],i[1],\cdots,i[M-1]]^{\mathrm{T}},\\
		&\mathbf{n}  \!=\![n_{r}[0],n_{r}[1],\cdots,n_{r}[M-1]]^{\mathrm{T}}.
	\end{align}
\end{subequations}
Here, element $e^{j2\pi f_{k}^{\mathrm{d}}\frac{nT_{{\rm sen}}}{M}}$
of diagonal matrix $\mathbf{D}(v_{k})$ represents the intrapulse Doppler effect, and vector $\mathbf{d}_{\mathrm{idft},m_{k}}$ is the $m_{k}$-th column of the inverse DFT (IDFT) matrix of dimension $M$. The user parameters of interest are included in $\mathbf{D}(v_{k})$, 
$\tilde{S}_{m_{k}}(\tau_{k})$, and $\mathbf{d}_{\mathrm{idft},m_{k}}$.

A given user reflects signals at different frequencies, each with a different reflection gain. The frequency yielding the highest reflection gain can be regarded as the central reflection frequency
of the $k$-th user, denoted as $f_{k_{\mathrm{c}}}$, where  $k_{\mathrm{c}}\in\mathcal{M}_{k}$.
The angle corresponding to the main lobe peak is regarded as the estimated angle of the $k$-th user. 
In the subsequent radar detection process, the first step is to segregate the signal components at  different frequencies within the echo signal in
(\ref{eqjii0}). Thus, we determine the user angles based on the frequency index and proceed to estimate the users' distance
and velocity parameters from the isolated subcarrier-specific echo
signals. To this end, we define \begin{align}\tilde{\mathbf{d}}_{\mathrm{idft},m_{k}}(v_{k})  =\mathbf{D}(v_{k})\mathbf{d}_{\mathrm{idft},m_{k}},\label{xe45}
\end{align} 
and rewrite (\ref{eqjii0}) as 
\begin{align}
	\mathbf{y} & =\mathbf{D}_{M}\sum_{k=1}^{K}\sum_{m_{k}\in\mathcal{M}_{k}}\tilde{\mathbf{d}}_{\mathrm{idft},m_{k}}(v_{k})\tilde{S}_{m_{k}}(\tau_{k})+\mathbf{i}+\mathbf{n}.\label{eq:7y8u}
\end{align}

\textbf{(1.1) Single-Antenna Receiver for Low Doppler Frequencies}

Under the condition of low Doppler frequencies, bounded by half of the subcarrier spacing, i.e., $|f_{k}^{\mathrm{d}}|<\triangle f_{{\rm sen}}/2$,
resulting in a user velocity constraint of $|v_{k}|<\frac{c\triangle f_{{\rm sen}}}{4f_{\mathrm{c}}}$,
DFT can be used to estimate the central reflection frequency associated with each user \cite{2011radarproc}. In particular, by taking the DFT of $\mathbf{y}$, one obtains 
\begin{align}
	\mathbf{y}_{\mathrm{DFT}}  &=\mathbf{D}_{\mathrm{dft}}\mathbf{D}_{M}^{-1}\mathbf{y}\nonumber \\
	& =\mathbf{D}_{\mathrm{dft}}\!\!\left(\!\sum_{k=1}^{K}\!\sum_{m_{k}\in\mathcal{M}_{k}}\!\!\!\!\tilde{\mathbf{d}}_{\mathrm{idft},m_{k}}(v_{k})\tilde{S}_{m_{k}}(\tau_{k})\!+\!\mathbf{D}_{M}^{-1}(\mathbf{i}\!+\!\mathbf{n})\right)\!,
\end{align}
where the dimension of DFT matrix $\mathbf{D}_{\mathrm{dft}}$ is
$M$, and the noise-free version of vector $\mathbf{y}_{\mathrm{DFT}}$
contains $\sum_{k=1}^{K}|\mathcal{M}_{k}|$ nonzero values at row indices $\bigcup_{k=1}^{K}\mathcal{M}_{k}$. Since the subcarrier representing the central reflection frequency $f_{k_{\mathrm{c}}}$ has a higher reflection gain compared to the other subcarriers in $\mathcal{M}_{k}$, a one-dimensional search can be effectively applied to identify the $K$ row indices correspongding to power peaks in $\mathbf{y}_{\mathrm{DFT}}$.  These  $K$ row indices  serve as estimates for the  indices of the central reflection frequencies, and the  corresponding angles  in the codebook of the rainbow beams serve as  estimates of the user direction angles, denoted by $\hat{\theta}_{k}$.  Note that the intrapulse Doppler effect $e^{j2\pi f_{k}^{\mathrm{d}}\frac{nT_{{\rm sen}}}{M}}$ does not cause an estimation error for  the central reflection frequencies, as the subcarriers  are not shifted to  adjacent subcarrier frequencies  owing to the assumption of a low Doppler frequency, i.e.,  $|f_{k}^{\mathrm{d}}|<\triangle f_{{\rm sen}}/2$. However, errors in user angle estimation may arise due to a mismatch between the discrete angles in the codebook of the rainbow beam and the continuous user angles. This error can be reduced by increasing the number of subcarriers.

Based on the estimate of $f_{k_{\mathrm{c}}}$, denoted by $f_{\hat{k}_{\mathrm{c}}}$, the Doppler frequencies and delays can be  estimated by formulating the following problem: 
\begin{equation}
	\min_{\{v_{k},\mathbf{s}_{k}(\tau_{k})\}_{k=1}^{K}}\left\Vert \mathbf{D}_{M}^{-1}\mathbf{y}-\sum_{k=1}^{K}\mathbf{D}(v_{k})\mathbf{D}_{\mathrm{idft},k}\mathbf{s}_{k}(\tau_{k})\right\Vert ,\label{eq:r35}
\end{equation}
where matrix $\mathbf{D}_{\mathrm{idft},k}\in\mathbb{C}^{M\times|\mathcal{M}_{k}|}$
has $\mathbf{d}_{\mathrm{idft},m_{k}},m_{k}\in\mathcal{M}_{k}$, in (\ref{b4}) as its columns in ascending order of $m_{k}$ and vector $\mathbf{s}_{k}(\tau_{k})\in\mathbb{C}^{|\mathcal{M}_{k}|\times1}$ contains the corresponding $\tilde{S}_{m_{k}}(\tau_{k}),m_{k}\in\mathcal{M}_{k}$,
given in (\ref{b3}) as elements  in decending order of $m_{k}$. 

In order to solve problem (\ref{eq:r35}) efficiently, we   assume that the interference between the users is weak, i.e., non-overlapping sets $\mathcal{M}_{k_{1}}\cap\mathcal{M}_{k_{2}}\neq\emptyset$
for $k_{1}\neq k_{2}$\footnote{This angular resolution assumption requires that adjacent users are separated by at least 	one mainlobe width, i.e., $4\pi/N_{t}$ \cite[Ch. 22.7]{EW-antenna}. 	For $N_{t}=256$, the angular resolution between users is $2.8^{\circ}$.}.  Assuming further a large $M$, we obtain the approximation  $\mathbf{D}_{\mathrm{idft},k_{1}}^{\mathrm{H}}\mathbf{D}^{\mathrm{H}}(v_{k_{1}})\mathbf{D}(v_{k_{2}})\mathbf{D}_{\mathrm{idft},k_{2}}\approx\mathbf{I}$ for $k_{1}\neq k_{2}$, where $\mathbf{I}$ is the identity matrix. Given $\{v_{k}\}_{k=1}^{K}$, the  optimal estimate for $\{\mathbf{s}_{k}(\tau_{k})\}_{k=1}^{K}$ based on (\ref{eq:r35}) is obtained as 
\begin{equation}
\hat{\mathbf{s}}_{k}(\tau_{k})=\left(\mathbf{D}_{\mathrm{idft},k}^{\mathrm{H}}\mathbf{D}_{\mathrm{idft},k}\right)^{-1}\mathbf{D}_{\mathrm{idft},k}^{\mathrm{H}}\mathbf{D}^{\mathrm{H}}(v_{k})\mathbf{D}_{M}^{-1}\mathbf{y},\forall k.\label{eq:f45-2}
\end{equation}
Inserting (\ref{eq:f45-2}) into Problem (\ref{eq:r35}), the  $\{v_{k}\}_{k=1}^{K}$ can be obtained as 
\begin{equation}
	\{\hat{v}_{k}\}_{k=1}^{K}=\mathrm{arg}\min_{\{v_{k}\}_{k=1}^{K}}\sum_{k=1}^{K}||\mathbf{y}^{H}\left(\mathbf{D}_{M}^{\mathrm{H}}\right)^{-1}\mathbf{D}(v_{k})\mathbf{D}_{\mathrm{idft},k}||^2.\label{eq:r35-1}
\end{equation}
Conveniently Problem (\ref{eq:r35-1}) can be decomposed into $K$ independent subproblems, i.e.,  \\
$\min_{v_{k}}||\mathbf{y}^{H}\left(\mathbf{D}_{M}^{\mathrm{H}}\right)^{-1}\mathbf{D}(v_{k})\mathbf{D}_{\mathrm{idft},k}||^2, \forall k$,  which can  be  solved through a one-dimensional search in  interval $v_{k}\in[-\frac{1}{2},\frac{1}{2}]\frac{c}{2f_{\mathrm{c}}T_{{\rm sen}}}$ in parallel.

Next, for estimation of the delay, from (\ref{b3}) we obain the following relationship
\begin{align}
	e^{j2\pi\triangle f_{{\rm sen}}\tau_{k}} & =\frac{\tilde{S}_{m_{k}}(\tau_{k})\left/\left(\mathbf{a}_{N_{t}}^{\mathrm{H}}\left(\theta_{k},f_{\mathrm{c}}+f_{m_{k}}\right)\mathbf{f}_{m_{k}}S_{m_{k}}\right)\right.}{\tilde{S}_{m_{k}+1}(\tau_{k})\left/\left(\mathbf{a}_{N_{t}}^{\mathrm{H}}\left(\theta_{k},f_{\mathrm{c}}+f_{m_{k}+1}\right)\mathbf{f}_{m_{k}+1}S_{m_{k}+1}\right)\right.},\label{eq:ji9}
\end{align}
which implies that the delay estimate, denoted as $\hat{\tau}_{k}$, can be readily obtained once the scalar values $\tilde{S}_{m_{k}}(\tau_{k})$ for  two subcarriers from set $\mathcal{M}_{k}$ are known. To maximize estimation accuracy, it is natural to select the two subcarriers with
the highest reflection gains from $\mathcal{M}_{k}$, namely the central reflection frequency $f_{k_{\mathrm{c}}}$ and an adjacent subcarrier
($f_{k_{\mathrm{c}}+1}$ or $f_{k_{\mathrm{c}}-1}$) denoted as $f_{k_{\mathrm{c}}\pm1}$.
Then,  according to  (\ref{eq:f45-2}), we have 
\begin{subequations}
	\label{hc-1} 
	\begin{align}
		\hat{\tilde{S}}_{\hat{k}_{\mathrm{c}}}(\tau_{k}) & =\frac{1}{M}[\mathbf{D}_{\mathrm{dft}}]_{\hat{k}_{\mathrm{c}},:}\mathbf{D}^{\mathrm{H}}(\hat{v}_{k})\mathbf{D}_{M}^{-1}\mathbf{y},\label{s1}\\
		\hat{\tilde{S}}_{\hat{k}_{\mathrm{c}}\pm1}(\tau_{k}) & =\frac{1}{M}[\mathbf{D}_{\mathrm{dft}}]_{\hat{k}_{\mathrm{c}}\pm1,:}\mathbf{D}^{\mathrm{H}}(\hat{v}_{k})\mathbf{D}_{M}^{-1}\mathbf{y},\label{eq:s2}
	\end{align}
\end{subequations}
due to the orthogonality of OFDM subcarriers. Based on 
(\ref{eq:ji9}) and (\ref{hc-1}), we obtain the estimate 
\begin{align}
	\hat{\tau}_{k} & =\pm\frac{1}{2\pi\triangle f_{{\rm sen}}}\angle\left(\frac{\hat{\tilde{S}}_{\hat{k}_{\mathrm{c}}}(\tau_{k})\left/\left(N_{t}S_{\hat{k}_{\mathrm{c}}}\right)\right.}{\hat{\tilde{S}}_{\hat{k}_{\mathrm{c}}\pm1}(\tau_{k})\left/\left(\mathbf{a}_{N_{t}}^{\mathrm{H}}\left(\hat{\theta}_{k},f_{\mathrm{c}}+f_{\hat{k}_{\mathrm{c}}\pm1}\right)\mathbf{f}_{\hat{k}_{\mathrm{c}}\pm1}S_{\hat{k}_{\mathrm{c}}\pm1}\right)\right.}\right),\label{eq:ji9-1}
\end{align}
where we exploited  $N_{t}=\mathbf{a}_{N_{t}}^{\mathrm{H}}\left(\hat{\theta}_{k},f_{\mathrm{c}}+f_{\hat{k}_{\mathrm{c}}}\right)\mathbf{f}_{\hat{k}_{\mathrm{c}}}$ since precoding vector $\mathbf{f}_{\hat{k}_{\mathrm{c}}}$ is designed to align with the steering vector in direction $\hat{\theta}_{k}$.

\textbf{(1.2) Single-Antenna Receiver for High Doppler Frequencies}\textbf{}

In scenarios involving high-speed movement, Doppler ambiguity occurs when the absolute value of the Doppler frequency is greater than half of the subcarrier spacing, i.e., $|f_{k}^{\mathrm{d}}|\geq\triangle f_{{\rm sen}}/2$. This Doppler ambiguity affects  $\tilde{\mathbf{d}}_{\mathrm{idft},m_{k}}(v_{k})$ in (\ref{xe45}). To elaborate, let $f_{k}^{\mathrm{d}}T_{{\rm sen}}$ be divided into $f_{k}^{\mathrm{d}}T_{{\rm sen}}=\varepsilon_{\mathrm{i},k}+\varepsilon_{\mathrm{f},k}$, where $\varepsilon_{\mathrm{i},k}$ is the integer component and $\varepsilon_{\mathrm{f},k}$
is the fractional component such that $-1/2\leq\varepsilon_{\mathrm{f},k}<1/2$.
$\varepsilon_{\mathrm{i},k}$ shifts the frequency $f_{m_{k}}$ to $f_{m_{k}+\varepsilon_{\mathrm{i},k}}$ at the receiver. This may introduce  Doppler ambiguity if the shift, $\varepsilon_{\mathrm{i},k}$, is not correctly estimated.  The DFT-based method proposed in the previous subsection cannot resolve this  Doppler ambiguity.  To tackle this issue, an alternative method exploiting   two OFDM symbols is proposed, where the Doppler frequency can be estimated by utilizing the interpulse Doppler effect.

Define the OFDM block duration as $T_{\mathrm{sym}}=T_{\mathrm{cp}}+T_{{\rm sen}}$,
where $T_{\mathrm{cp}}$ is  the CP duration, which limits the maximum detection distance, denoted as $R_{\mathrm{max}}$, as
$\frac{2R_{\mathrm{max}}}{c}\leq T_{\mathrm{cp}}$.  The first OFDM symbol is given by
$\sum_{m \in \mathcal{M}}s_{m}(t)$  as before. The  second  OFDM symbol is expressed as $\sum_{m \in \mathcal{M}}s_{m}^{(2)}(t-T_{\mathrm{sym}})\triangleq  \sum_{m \in \mathcal{M}}S_{m}^{(2)}e^{j2\pi f_{m}(t-T_{\mathrm{sym}})}$, where $S_{m}^{(2)}$ is the pilot symbol on the $m$-th subcarrier of the second OFDM symbol.  Following the same signal processing steps as for the first OFDM
symbol and removing the CP, for a single-antenna receiver, the received down-converted baseband signal corresponding to (\ref{eq:ji45}) is given by  
$	y^{(2)}(t) 
	=\sum_{k=1}^{K}\left(\sum_{m_{k}\in\mathcal{M}_{k}}z^{(2)}_{m_{k}}(t)+\sum_{m_{k}\in\mathcal{M}\setminus\mathcal{M}_{k}} z^{(2)}_{m_{k}}(t)\right)+n^{(2)}(t)$,
where  $z^{(2)}_{m_{k}}(t)=\sqrt{P_{{\rm s}}}\tilde{\beta}_{k}\mathbf{a}_{N_{t}}^{\mathrm{H}}\left(\theta_{k},f_{\mathrm{c}}+f_{m_{k}}\right)\mathbf{f}_{m_{k}}e^{j2\pi f_{k}^{\mathrm{d}}t}s^{(2)}_{m_{k}}\left(t-T_{\mathrm{sym}}-\tau_{k}\right)$ and $n^{(2)}(t)$ is the AWGN.

We sample the matched filtered $y^{(2)}(t)$ at times $t=T_{\mathrm{sym}}+\frac{nT_{{\rm sen}}}{M}$, 
and obtain the discrete-time  signal as 
\begin{align}
	y^{(2)}[n] 
	=\sum_{k=1}^{K}\left(\sum_{m_{k}\in\mathcal{M}_{k}}z^{(2)}_{m_{k}}[n]+\sum_{m_{k}\in\mathcal{M}\setminus\mathcal{M}_{k}} z^{(2)}_{m_{k}}[n]\right)+n^{(2)}[n],\label{eq:bj87-1}
\end{align}
where $z^{(2)}_{m_{k}}[n]
=\sqrt{P_{{\rm s}}}\tilde{\beta}_{k}\mathbf{a}_{N_{t}}^{\mathrm{H}}\left(\theta_{k},f_{\mathrm{c}}+f_{m_{k}}\right)\mathbf{f}_{m_{k}}e^{j2\pi f_{k}^{\mathrm{d}}\frac{nT_{{\rm sen}}}{M}}e^{j2\pi f_{k}^{\mathrm{d}}T_{\mathrm{sym}}}S_{m_{k}}^{(2)}e^{-j2\pi f_{m_{k}}\tau_{k}}e^{j2\pi m_{k}\frac{n}{M}}$ and \\
$e^{j2\pi f_{k}^{\mathrm{d}}T_{\mathrm{sym}}}$ represents  the interpulse
Doppler effect. Furthermore, for the second OFDM symbol, (\ref{eq:7y8u}) becomes 
\begin{align}
\mathbf{y}^{(2)} & =\mathbf{D}_{M}\sum_{k=1}^{K}\sum_{m_{k}\in\mathcal{M}_{k}}\tilde{\mathbf{d}}_{\mathrm{idft},m_{k}}(v_{k})\tilde{S}_{m_{k}}^{(2)}(\tau_{k},v_{k})+\mathbf{i}^{(2)}+\mathbf{n}^{(2)},\label{eq:7y8u-1}
\end{align}
where $\tilde{S}_{m_{k}}^{(2)}(\tau_{k},v_{k})=\sqrt{P_{{\rm s}}}\tilde{\beta}_{k}\mathbf{a}_{N_{t}}^{\mathrm{H}}\left(\theta_{k},f_{\mathrm{c}}+f_{m_{k}}\right)\mathbf{f}_{m_{k}}S_{m_{k}}^{(2)}e^{-j2\pi f_{m_{k}}\tau_{k}}e^{j2\pi f_{k}^{\mathrm{d}}T_{\mathrm{sym}}}$,
and $\mathbf{i}^{(2)}$ and $\mathbf{n}^{(2)}$ are the interference
from the sidelobes and the AWGN, respectively. Now,  the Doppler frequency
can be obtained by comparing the phase difference between $\tilde{S}_{m_{k}}(\tau_{k})$
in (\ref{b3}) and $\tilde{S}_{m_{k}}^{(2)}(\tau_{k},v_{k})$, given
by 
\begin{align}
e^{j2\pi f_{k}^{\mathrm{d}}T_{\mathrm{sym}}} & =\frac{\tilde{S}_{m_{k}}^{(2)}(\tau_{k},v_{k})/S_{m_{k}}^{(2)}}{\tilde{S}_{m_{k}}(\tau_{k})/S_{m_{k}}}.\label{eq:2f4}
\end{align}

Therefore, for the first and second OFDM symbols, the DFT and phase
rotation techniques are used to estimate $k_{\mathrm{c}}+\varepsilon_{\mathrm{i},k}$
and $\varepsilon_{\mathrm{f},k}$, respectively. Defining $\hat{k}_{\mathrm{c+i}}$
and $\hat{\varepsilon}_{\mathrm{f},k}$ as the estimates of  $k_{\mathrm{c}}+\varepsilon_{\mathrm{i},k}$
and $\varepsilon_{\mathrm{f},k}$, respectively, and following (\ref{s1}),
we obtain the estimates 
\begin{subequations}
\label{hc-4} 
\begin{align}
\hat{\tilde{S}}_{\hat{k}_{\mathrm{c+i}}}(\tau_{k}) & =\frac{1}{M}[\mathbf{D}_{\mathrm{dft}}]_{\hat{k}_{\mathrm{c+i}},:}\mathbf{D}^{\mathrm{H}}\left(\frac{c\hat{\varepsilon}_{\mathrm{f},k}}{2f_{\mathrm{c}}T_{{\rm sen}}}\right)\mathbf{D}_{M}^{-1}\mathbf{y},\label{s1-1}\\
\hat{\tilde{S}}_{\hat{k}_{\mathrm{c+i}}}^{(2)}(\tau_{k},v_{k}) & =\frac{1}{M}[\mathbf{D}_{\mathrm{dft}}]_{\hat{k}_{\mathrm{c+i}},:}\mathbf{D}^{\mathrm{H}}\left(\frac{c\hat{\varepsilon}_{\mathrm{f},k}}{2f_{\mathrm{c}}T_{{\rm sen}}}\right)\mathbf{D}_{M}^{-1}\mathbf{y}^{(2)}.\label{eq:s2-1}
\end{align}
\end{subequations}

According to (\ref{eq:2f4}) and utilizing (\ref{hc-4}),
we have 
\begin{align}
\hat{f}_{k}^{\mathrm{temp}} & =\frac{1}{2\pi T_{\mathrm{sym}}}\angle\left(\frac{\hat{\tilde{S}}_{\hat{k}_{\mathrm{c+i}}}^{(2)}(\tau_{k},v_{k})\left/S_{\hat{k}_{\mathrm{c+i}}}^{(2)}\right.}{\hat{\tilde{S}}_{\hat{k}_{\mathrm{c+i}}}(\tau_{k})\left/S_{\hat{k}_{\mathrm{c+i}}}\right.}\right).\label{eq:ji9-1-1}
\end{align}

Note that $e^{j2\pi f_{k}^{\mathrm{d}}T_{\mathrm{sym}}}$ is a periodic function of $f_{k}^{\mathrm{d}}T_{\mathrm{sym}}$ with period 1, {so $\hat{f}_{k}^{\mathrm{temp}}$ may not be the estimate of $f_{k}^{\mathrm{d}}$.} The estimate for the Doppler frequency, denoted by $\hat{f}_{k}^{\mathrm{d}}$, is obtained based on the following analysis. 1) For $0\leq f_{k}^{\mathrm{d}}T_{\mathrm{sym}}=\varepsilon_{\mathrm{f},k}<1/2$, 
resulting in estimates  $\hat{f}_{k}^{\mathrm{temp}}T_{\mathrm{sym}}>0$
and $\hat{\varepsilon}_{\mathrm{f},k}>0$ with $\left|\hat{f}_{k}^{\mathrm{d}}T_{\mathrm{sym}}-\hat{\varepsilon}_{\mathrm{f},k}\right| \rightarrow 0$, then we have $\hat{f}_{k}^{\mathrm{d}}=\hat{f}_{k}^{\mathrm{temp}}$.
2) For $1/2\leq f_{k}^{\mathrm{d}}T_{\mathrm{sym}}=1+\varepsilon_{\mathrm{f},k}<1$, we obtain estimates
$\hat{f}_{k}^{\mathrm{temp}}T_{\mathrm{sym}}<0$ and $-1/2<\hat{\varepsilon}_{\mathrm{f},k}<0$, but $\hat{\varepsilon}_{\mathrm{f},k}$ is inaccurate due to the failure of angle rotation method at high Doppler frequency,  resulting in $\left|\hat{f}_{k}^{\mathrm{d}}T_{\mathrm{sym}}-\hat{\varepsilon}_{\mathrm{f},k}\right| \gg 0$. Hence, we obtain $\hat{f}_{k}^{\mathrm{d}}T_{\mathrm{sym}}=\hat{f}_{k}^{\mathrm{temp}}T_{\mathrm{sym}}+1$.
3) For  $1\leq f_{k}^{\mathrm{d}}T_{\mathrm{sym}}=1+\varepsilon_{\mathrm{f},k}<3/2$, we acquire estimate $\hat{f}_{k}^{\mathrm{d}}T_{\mathrm{sym}}>0$ and inaccurate estimate $1/2>\hat{\varepsilon}_{\mathrm{f},k}>0$, leading to  $\left|\hat{f}_{k}^{\mathrm{d}}T_{\mathrm{sym}}-\hat{\varepsilon}_{\mathrm{f},k}\right| \gg 0$, then we have $\hat{f}_{k}^{\mathrm{d}}T_{\mathrm{sym}}=\hat{f}_{k}^{\mathrm{temp}}T_{\mathrm{sym}}+1$.
Therefore, the esitmate $\hat{f}_{k}^{\mathrm{d}}$ can be summarized as
\begin{equation}
\hat{f}_{k}^{\mathrm{d}}=\begin{cases}
\hat{f}_{k}^{\mathrm{temp}}, & \textrm{if }\left|\hat{f}_{k}^{\mathrm{d}}T_{\mathrm{sym}}-\hat{\varepsilon}_{\mathrm{f},k}\right|<\epsilon,\\
\frac{\hat{f}_{k}^{\mathrm{temp}}T_{\mathrm{sym}}+1}{T_{\mathrm{sym}}}, & \textrm{if }\left|\hat{f}_{k}^{\mathrm{d}}T_{\mathrm{sym}}-\hat{\varepsilon}_{\mathrm{f},k}\right|\geq\epsilon,
\end{cases}\label{eq:ii0}
\end{equation}
where $\epsilon$ is a threshold and can be empirically  set to 0.05. For example, assuming a system with
$f_{\mathrm{c}}=100$ GHz, $\triangle f_{{\rm sen}}=240$
kHz, and $T_{\mathrm{sym}}=\frac{5}{4}T_{{\rm sen}}$, the  maximum velocity that can be estimated based on (\ref{eq:ii0})
is 432 m/s,  which is sufficient for all civilian mobile devices. The maximum velocity can be further increased by using a lower central frequency or a larger subcarrier spacing.

The index of the central reflection frequency for the  $k$-th user can be obtained as 
\begin{align}
\hat{k}_{\mathrm{c}} & =\left\lfloor \hat{k}_{\mathrm{c+i}}+\hat{\varepsilon}_{\mathrm{f},k}-\hat{f}_{k}^{\mathrm{d}}\right\rceil ,
\end{align}
where the operation $\left\lfloor \cdot\right\rceil $ rounds towards the nearest integer, which ensures that   $\hat{k}_{\mathrm{c}}$ is not affected by the inaccurate estimate $|\hat{\varepsilon}_{\mathrm{f},k}|<1/2$. Once the estimate $\hat{k}_{\mathrm{c}}$ has been determined,  the corresponding angles can be obtained from the rainbow beam  codebook and the resulting delays can be derived from (\ref{eq:s2}) and (\ref{eq:ji9-1}).

\textbf{(2) Multi-Antenna Receiver}

\label{MA1O}

Given the fleeting nature of real-time location data for high-speed  mobile users,   the previous approach, exploiting two OFDM symbols for training,  is suboptimal. 
To reduce the training time overhead inherent to user detection, a multi-antenna receiver is considered as it introduces additional degrees of freedom for estimation. 
In particular, the user angle can be estimated first by exploiting the diversity gain introduced by  multiple antennas\footnote{Although,  for the multi-antenna receiver, the angle of the user is not directly estimated based on the subcarrier frequency,  rainbow beams are  still beneficial as their  multiple beams simultaneously  cover the entire angular  space for fast beam training.}, thereby determining the central reflection frequency, $\mathbf{f}_{m_{k}}$.   Then, $\varepsilon_{\mathrm{i},k}$ can be  acquired once $\mathbf{f}_{m_{k}+\varepsilon_{\mathrm{i},k}}$ has been obtained.

By defining $\boldsymbol{\theta}=[\theta_{1},...,\theta_{K}]^{\mathrm{T}}$ and $r_k[n]=\sum_{m_{k}\in\mathcal{M}_{k}}z_{m_{k}}[n]+\sum_{m_{k}\in\mathcal{M}\setminus\mathcal{M}_{k}} z_{m_{k}}[n]$, we rewrite  (\ref{eq:jt6t65}) for the multi-antenna receiver  in  compact form as follows
\begin{align}\mathbf{y}[n]  =\mathbf{A}_{N_{r}}(\boldsymbol{\theta})\mathbf{r}[n]+\mathbf{n}_{r}[n],\label{eq:ji8l-1}
\end{align}
where $\mathbf{A}_{N_{r}}(\boldsymbol{\theta})=[\mathbf{a}_{N_{r}}(\theta_{1}),\ldots,\mathbf{a}_{N_{r}}(\theta_{K})]$
and $\mathbf{r}[n]=[r_{1}[n],\ldots,r_{K}[n]]^{\mathrm{T}}$.
Determining $\boldsymbol{\theta}$ from (\ref{eq:ji8l-1}) is a typical
angle estimation problem \cite{lee-proceeding}, which can be solved through either grid-on estimation methods (including DFT and orthogonal matching pursuit (OMP) \cite{OMP-TIT}), characterized by low complexity, or gridless estimation methods (including multiple signal classification (MUSIC) \cite{music} and atomic norm minimization (ANM) \cite{anm-2013}),
known for their high precision. The grid-on estimation methods yield
sufficient estimation accuracy for large $N_{r}$. However, considering the limitation of a small $N_{r}$, which can reduce hardware cost and avert the beam squint effect at the receiving array, we adopt the high-precision ANM method to estimate the angles. Notably, the ANM method's complexity, which is $\mathcal{O}(N_{r}^{3})$, is acceptable for small $N_{r}$.

We collect $M$ vectors $\mathbf{y}_{r}[n]$ in matrix 
$\mathbf{Y}_{r}=[\mathbf{y}_{r}[0],...,\mathbf{y}_{r}[M-1]]$,  whose
rank is $K$ as $K<N_{r}$ and $K<M$.  Then, by  constructing a matrix
$\mathbf{D}_{Y,1}\in\mathbb{C}^{M\times K}$ comprising the right
singular vectors corresponding to the $K$ non-zero singular values
of $\mathbf{Y}_{r}$, the ANM optimization problem can be  formulated as
follows \cite{anm-2013} 
\begin{subequations}
	\label{Pro:min-power-f} 
	\begin{align}
		\{\hat{\mathbf{U}}_{s},\hat{\mathbf{Z}}_{s},\hat{\mathbf{z}}\}=\mathrm{arg}\min_{\mathbf{U}_{s},\mathbf{Z}_{s},\mathbf{z}} & \;\frac{\nu}{2}\left(\mathrm{Tr}(\mathbf{U}_{s})+\mathrm{Tr}(\mathrm{Toe}(\mathbf{z}))\right)+\frac{1}{2}||\mathbf{Y}_{r}\mathbf{D}_{Y,1}-\mathbf{Z}_{s}||_{F}^{2}\label{eq:F93}\\
		\mathrm{s.t.} & \;\left[\begin{array}{cc}
			\mathbf{U}_{s} & \mathbf{Z}_{s}^{\mathrm{H}}\\
			\mathbf{Z}_{s} & \mathrm{Toe}(\mathbf{z})
		\end{array}\right]\succeq0,\label{eq:444g}
	\end{align}
\end{subequations}
where regularization parameter $\nu\propto\sqrt{\sigma^{2}_{\rm s}N_{r}\log(N_{r})}$
controls the trade-off between promoting sparsity and ensuring data
fitting, and $\mathbf{Z}_{s}=\mathbf{Z}_{0}\mathbf{D}_{Y,1}$ with sparse
matrix $\mathbf{Z}_{0}\in\mathbb{C}^{N_{r}\times M}$, which is a linear
combination of a few $\mathbf{a}_{N_{r}}(\theta)$. The Hermitian Toeplitz matrix $\mathrm{Toe}(\mathbf{z})\in\mathbb{C}^{N_{r}\times N_{r}}$ is characterized by its first row  $\mathbf{z}\in\mathbb{C}^{N_{r}\times1}$, effectively collecting the $\mathbf{a}_{N_{r}}(\theta)$ used to construct $\mathbf{Z}_{0}$. Thus, the estimate $\hat{\boldsymbol{\theta}}$ can be obained with  a root-finding approach, e.g.,
root MUSIC \cite{music}, based on  $\mathrm{Toe}(\hat{\mathbf{z}})$.

Upon having obtained an  accurate estimate $\hat{\boldsymbol{\theta}}$,
the primary objective is to identify the discrete angles of  the rainbow
codebook that best approximate $\hat{\boldsymbol{\theta}}$. Subsequently,
the subcarrier frequencies, denoted as $\{\hat{k}_{\mathrm{c}}\}_{k=1}^{K}$,
corresponding to the  identified discrete angles are regarded as estimates for the central reflection frequencies of the $K$ users.

Next, we estimate the Doppler frequencies. Given an accurate estimate $\hat{\boldsymbol{\theta}}$, the echo signal from the $k$-th
user can be extracted as follows
\begin{align}
	\mathbf{y}_{k} & =\left[\mathbf{A}_{N_{r}}^{\dagger}(\hat{\boldsymbol{\theta}})\mathbf{Y}_{r}\right]_{k,:}^{\mathrm{T}}\nonumber \\
	& =\left[r_{k}[0],...,r_{k}[M-1]\right]^{\mathrm{T}}+\mathbf{n}_{k}\nonumber \\
	& =\mathbf{D}_{M}\mathbf{D}(v_{k})\sum_{m_{k}\in\mathcal{M}_{k}}\mathbf{d}_{\mathrm{idft},m_{k}}\tilde{S}_{m_{k}}(\tau_{k})+\mathbf{i}_{k}+\mathbf{n}_{k},\label{eq:fe6}
\end{align}
where $\mathbf{n}_{k}$ is the corresponding  noise vector and $\mathbf{i}_{k}$ is the echo interefence caused by the side lobes of the remaining subcarriers reflected by the $k$-th user. 

Estimate $\hat{k}_{\mathrm{c+i}}$ can be easily acquired by finding the peak of the power  spectrum of $\mathbf{D}_{\mathrm{idft}}\mathbf{D}_{M+1}^{-1}\mathbf{y}_{k}$.
Furthermore, estimates of $\hat{\varepsilon}_{\mathrm{f},k}$ and $\tau_{k}$ can
be obtained by solving a simplified version of Problem (\ref{eq:r35}),   given by  
\begin{equation}
	\{\hat{v}_{\mathrm{f},k},\hat{\tau}_{k}\}=\mathrm{arg}\min_{v_{\mathrm{f},k},\tilde{\mathbf{s}}_{k}}\left\Vert \mathbf{D}_{M}^{-1}\mathbf{y}_{k}-\mathbf{D}(v_{\mathrm{f},k})\mathbf{D}_{\mathrm{idft},k}\mathbf{s}_{k}(\tau_{k})\right\Vert ,\label{eq:r35-2}
\end{equation}
where $\hat{\varepsilon}_{\mathrm{f},k}=\frac{2f_{\mathrm{c}}T_{{\rm sen}}}{c}\hat{v}_{\mathrm{f},k}$.
Finally, an estimate of the Doppler frequency is given by
\begin{align}
	\hat{f}_{k}^{\mathrm{d}} & =\hat{k}_{\mathrm{c+i}}+\hat{\varepsilon}_{\mathrm{f},k}-\hat{k}_{\mathrm{c}}.
\end{align}
Once  the estimate $\hat{f}_{k}^{\mathrm{d}}$ has been acquired,  the angles can be determined based on the rainbow codebook.

\subsection{Rainbow Beam Based User Tracking}

User tracking aims to swiftly update a user's new position for  the  subsequent frame, leveraging prior information regarding position and velocity. To support rapid user mobility, time overhead for position update has to be minimal, making low time overhead a crucial performance metric for user tracking schemes. Mathematically, let $\theta_{k}^{(i)}$
represent the physical direction of the $k$-th user in the $i$-th
frame. The objective of tracking is to quickly ascertain the physical
direction $\theta_{k}^{(i+1)}$ for the $(i+1)$-th frame based on
$\theta_{k}^{(i)}$. Utilizing prior information on the user's physical direction, distance, and velocity, the BS can predict the potential range of variation in the user's physical direction, denoted as  $[\theta_{k}^{(i)}-\alpha_{k}^{(i)},\theta_{k}^{(i)}+\alpha_{k}^{(i)}]$. Importantly, the variation distance $\alpha_{k}^{(i)}$
remains constant across many frames unless the user undergoes an abrupt
change in movement direction or velocity, constituting an irregular
motion. Therefore, the goal of the proposed tracking scheme
is to simultaneously establish multiple small beam sectors, encompassing the 
potential movement angles of the $K$ users. Based on the investigation
and analysis of rainbow beam training, we propose a simple and effective
strategy for user tracking. Leveraging the established rainbow codebook
designed for beam training across the entire angular range of $[0,\pi]$,
we first select subcarriers to generate multiple small beam sectors,
effectively covering the angular space relevant to tracking $K$
users. Specifically, let $\mathcal{M}_{k}^{\rm track}$ denote the set of subcarriers whose beams span the 
 angular interval $[\theta_{k}^{(i)}-\alpha_{k}^{(i)},\theta_{k}^{(i)}+\alpha_{k}^{(i)}]$.  Then, it is sufficient for the BS to transmit baseband pilot signal $\sum_{k=1}^{K}\sum_{m_{k}\in\mathcal{M}_{k}^{\rm  track}}S_{m}e^{j2\pi f_{m}t}$ to  track all users. The corresponding sampled received signal at the BS is given by 
\begin{align}
\mathbf{y}_{{\rm track}} & =\mathbf{D}_{M}\sum_{k=1}^{K}\sum_{m_{k}\in\mathcal{M}_{k}^{\rm  track}}\tilde{\mathbf{d}}_{\mathrm{idft},m_{k}}(v_{k})\tilde{S}_{m_{k}}(\tau_{k})+\mathbf{i}_{{\rm track}}+\mathbf{n},\label{eq:rakcin}
\end{align}
where $\mathbf{i}_{{\rm track}}$ contains the subcarrier interference from $\sum_{k^'=1, k^'\neq k}^{K}\mathcal{M}_{k^'}^{\rm  track}$, which is significantly
weaker compared to interference $\mathbf{i}$ caused by  $\mathcal{M}\setminus\mathcal{M}_{k}$ in (\ref{eq:7y8u}).
Based on the measurement in (\ref{eq:rakcin}) with weak interference, the Doppler frequency can be more accurately estimated using (\ref{eq:r35-1}).

In subsequent frames, as the angular spaces for user tracking  change,
the BS merely adapts the subcarrier frequency for tracking, eliminating
the need for reconfiguration of the precoding. This approach stands in contrast to existing user tracking methods, including the TTD-based user tracking scheme proposed in \cite{rainbow-linglong-tracking}, which maintains a  fixed pilot signal, while altering the tracking beam coverage through precoding adjustments involving PSs and TTDs.
In contrast, the proposed strategy tracks the varying beam coverage via the selection of different subcarriers and obviates precoding reconfiguration, thereby significantly mitigating hardware complexity and tracking-associated time overhead, which is beneficial for the realization of high-frequency devices.

\section{NUMERICAL RESULTS}
\label{simulation}

In this section, we evaluate the performance of the proposed rainbow
beam assisted ISAC system using the parameters specified in Table
I. All results are obtained by averaging over 1000 parameter realizations.
We consider a system with $f_{\mathrm{c}}=100$ GHz in our simulations.
In accordance with the 5G NR standard \cite{5gNR}, the available subcarrier
spacings are $15\times2^{n}$ kHz, where $n=\{1,2,3,4,5\}$. Most
subcarrier spacings support both data and synchronization signaling,
except for 60 kHz, dedicated to data transmission, and 240 kHz, which
is exclusively for synchronization signaling \cite{5gNR}. Therefore,
we choose $\triangle f_{{\rm sen}}=240$ kHz for sensing and $\triangle f_{\mathrm{com}}=120$
kHz for communication. The pilot symbols are set as $S_{m}=1,\forall m\in\mathcal{M}$, for sensing.   In addition, we define the sensing signal-to-noise ratio (SNR) for the considered   multiple-user scenario as $\mathrm{SNR}=P_{\rm s}\min\{|\beta_{k}|^{2},\forall k\}/\sigma_{\rm s}^{2}$, where $P_{\rm s}$ is the radar TX power, respectively.  To evaluate  performance, we calculate  the root mean squared errors (RMSEs) of the angle, distance, and velocity
estimates as $\frac{1}{K}\sum_{k=1}^{K}\mathbb{E}\left\{ |\theta_{k}-\hat{\theta}_{k}|^{2}\right\} $, $\frac{1}{K}\sum_{k=1}^{K}\mathbb{E}\left\{ |D_{k}-\hat{D}_{k}|^{2}\right\} $,
and $\frac{1}{K}\sum_{k=1}^{K}\mathbb{E}\left\{ |v_{k}-\hat{v}_{k}|^{2}\right\} $, respectively.

We compare the performance of the following schemes: 1) SA1O: The scheme relying on a  single-antenna receiver  and one set of rainbow beams proposed in Section \ref{SA1O} (1.1).  2) SA2O: The scheme employing  a  single-antenna receiver  and two sets of rainbow beams  proposed in Section \ref{SA1O} (1.2).  3) MA1O: The scheme proposed in Section \ref{MA1O} (2) for a  multi-antenna receiver.  4) SASC and MASC: Beam training via a single-carrier radar signal serves as  benchmarks. The radar signal is transmitted in $M$ time slots, directed towards different  angles within the domain $[0,\pi]$ to detect and estimate  users. Initially, a matched filter is employed to estimate the Doppler frequencies and delays, followed by the determination of the angles.  Single- and multi-antenna receivers are employed for SASC and MASC, respectively. More details can be found in \cite{2011radarproc}.

\begin{table*}
	\vspace{-1cm}
\centering \centering \caption{ISAC simulation parameters.}
\begin{tabular}{|>{\columncolor{mygray}}l|>{\columncolor{mygray}}l|l|l|}
\hline 
\textbf{Sensing} \textbf{Parameters} & \textbf{Value} & \cellcolor{mygreen} \textbf{Common Parameters} &\cellcolor{mygreen} \textbf{Value}\tabularnewline
\hline 
Number of RX Antennas, $N_{r}$  & 1 or 10 & \cellcolor{mygreen}Number of TX Antennas, $N_{t}$  & \cellcolor{mygreen}256 \tabularnewline
\hline 
Subcarrier Spacing, $\triangle f_{{\rm sen}}$  & 240 kHz  & \cellcolor{mygreen}Central Carrier Frequency, $f_{\mathrm{c}}$  & \cellcolor{mygreen}100 GHz \tabularnewline
\hline 
$\epsilon$-beamwidth,\textbf{ $\triangle\psi_{\epsilon}$}  & 0.886$\frac{2\pi}{N_{t}}$ & \cellcolor{mygreen}Frequency Guard Interval, $B_{\mathrm{guard}}$  & \cellcolor{mygreen}4.8 MHz\tabularnewline
\hline 
Number of Subcarriers, $M$  & 579 & \cellcolor{mygreen}Noise Power Density & \cellcolor{mygreen}$-174$ dBm/Hz\tabularnewline
\hline 
Symbol Duration, $T_{{\rm sen}}$  & 4.17 $\mu s$ & \cellcolor{mycyan}\textbf{Communication Parameters} &\cellcolor{mycyan} \tabularnewline
\hline 
Total Bandwidth, $B_{{\rm sen}}$  & 138.72 MHz & \cellcolor{mycyan}Subcarrier Spacing, $\triangle f_{\mathrm{com}}$  & \cellcolor{mycyan}120 kHz\tabularnewline
\hline 
Symbol Duration, $T_{\mathrm{cp}}$  & 0.52 $\mu s$ & \cellcolor{mycyan}Number of Subcarriers for Each User, $Q$  & \cellcolor{mycyan}1024\tabularnewline
\hline 
Total Symbol Duration, $T_{\mathrm{sym}}$  & 4.69 $\mu s$ & \cellcolor{mycyan}Total Bandwidth, $B_{{\rm com}}$  & \cellcolor{mycyan}122.88 MHz\tabularnewline
\hline 
Maximum Distance, $D_{\max}$  & 625 m  & \cellcolor{mygray}Maximum Velocity of SA2O, $v_{\max}^{\mathrm{SA2O}}$  & \cellcolor{mygray}480 m/s\tabularnewline
\hline 
Maximum Velocity of SA1O, $v_{\max}^{\mathrm{SA1O}}$  & 180 m/s  & \cellcolor{mygray}Maximum Velocity of MA1O, $Mv_{\max}^{\mathrm{SA1O}}$  & \cellcolor{mygray}104220 m/s \tabularnewline
\hline 
\end{tabular}
\vspace{-1cm}
\end{table*}

\subsection{Rainbow Beam User Training}

Due to the Doppler ambiguity, the performance of SA1O, SASC, and MASC is  studied in low-Doppler (LD)  scenarios, while SA2O and MA1O are evaluated for both LD and high-Doppler (HD)  scenarios.  The angles of the $K$ users are randomly generated in $[0,\pi]$, but  a minimum angle separation of at least $10{^\circ}$ between adjacent users is enforced.   The  distances and velocities of the $K$ users are randomly generated within given ranges, where the distance spans{\footnote{In the considered far-field scenario, the distance between BS and users must satisfy $R>r_{{\rm ray}}$, where $r_{{\rm ray}}=\frac{2D^{2}}{\lambda_{\mathrm{c}}}$ represents the Rayleigh distance. Here, $D$ denotes  the array aperture, given as $D=(N_{t}-1)d$ for a ULA. For the system parameters outlined	in Table I, the Rayleigh distance is calculated as $r_{{\rm ray}}=97.54$ m.}}  $[100,250]$ m, and the velocities  in the LD and HD scenarios span  $[10,150]$ m/s and $[180,450]$ m/s, respectively. 

\begin{figure}[htbp]
	\vspace{-0.7cm}
\centering \subfigure[Angle]{ %
\begin{minipage}[t]{0.315\linewidth}%
\centering \includegraphics[width=2.28in]{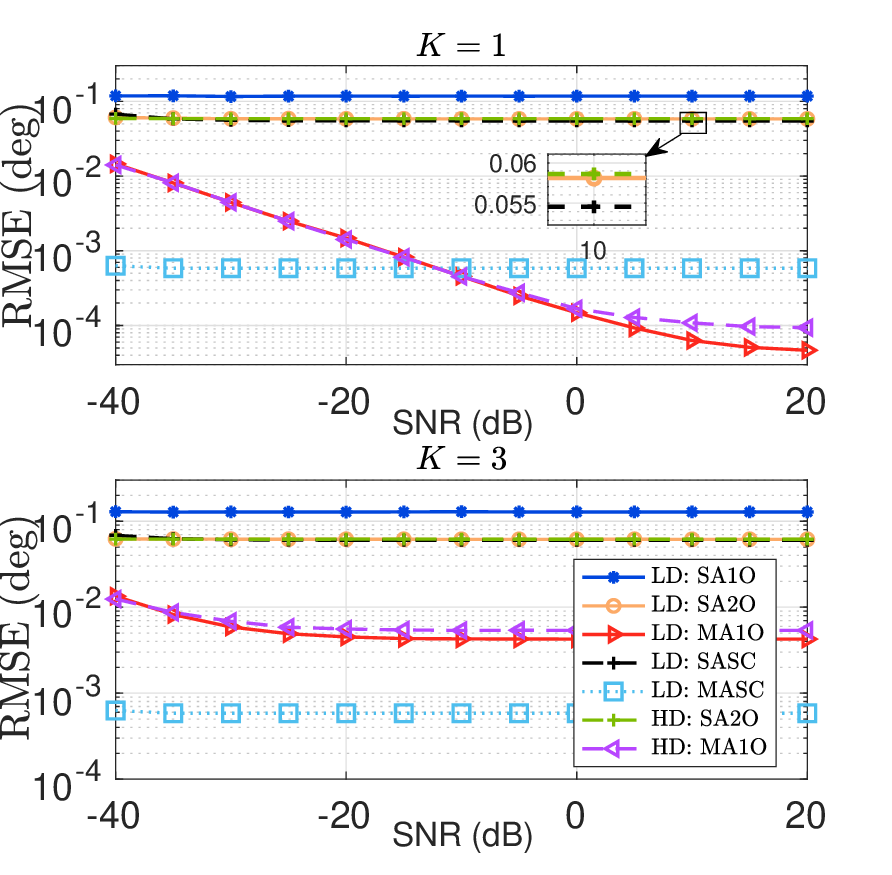}%
\end{minipage}} \subfigure[Velocity ]{ %
\begin{minipage}[t]{0.315\linewidth}%
\centering \includegraphics[width=2.28in]{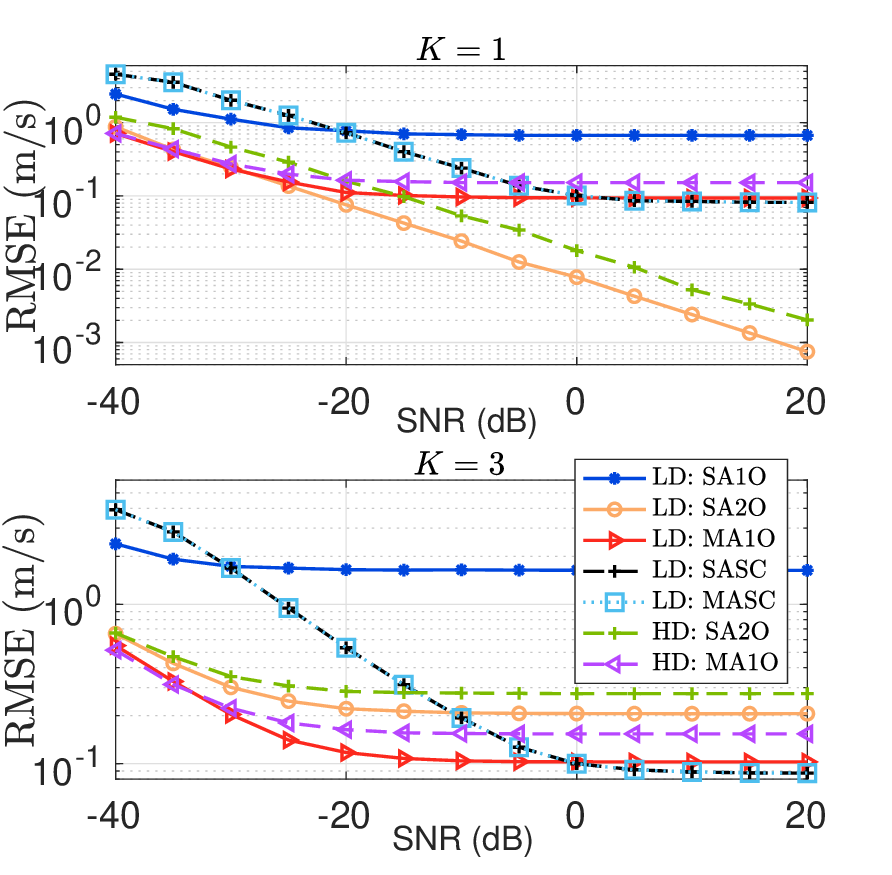}%
\end{minipage}}
\subfigure[Distance]{ %
\begin{minipage}[t]{0.315\linewidth}%
\centering \includegraphics[width=2.28in]{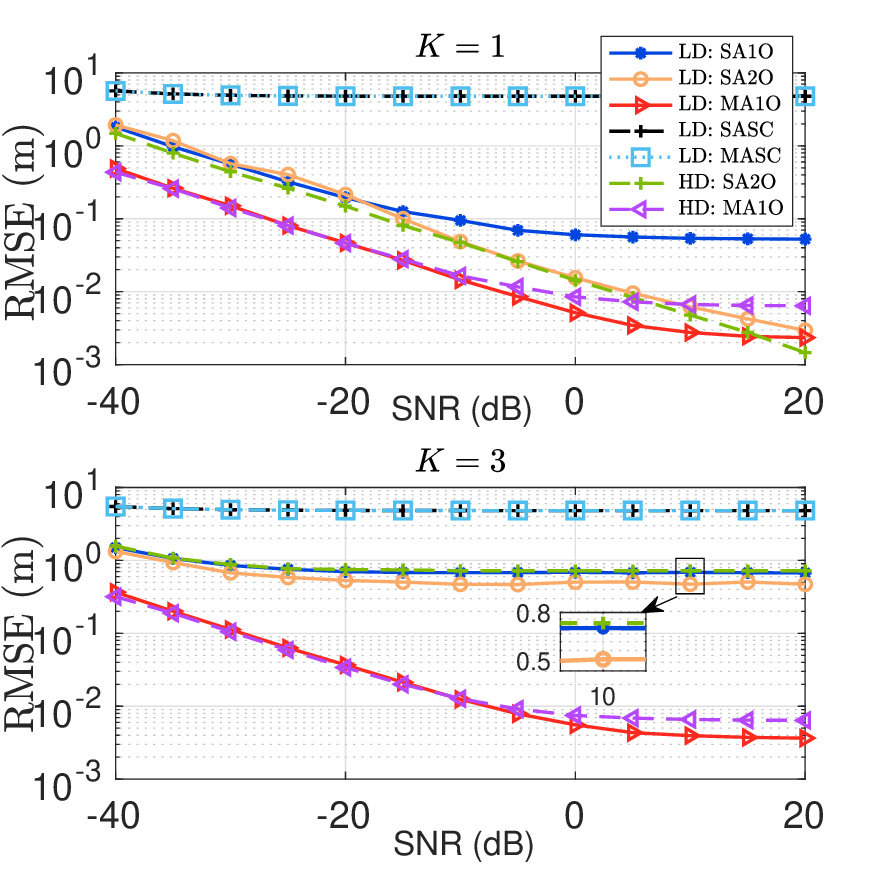}%
\end{minipage}} 
\vspace{-0.5cm}
\caption{Sensing performance versus SNR.}
\label{nmse_snr} 
\vspace{-0.8cm}
\end{figure}

Fig. \ref{nmse_snr} illustrates the estimation performance as a function of the sensing SNR for scenarios with $K=1$ and $K=3$ users. First,  as observed from Figs. \ref{nmse_snr}(a)-(c),     SA2O and MA1O  respectively yield  similar estimation performance for both low and high Doppler frequencies,   effectively handling the Doppler ambiguity while ensuring accuracy. The rainbow-beam approaches cause  inter-user-interference (IUI), which leads to  reduced estimation performance in multi-user scenarios, compared to single-user scenarios.  For further  analysis,  we focus on specific parameters in the multi-user scenario.  Regarding the angle estimation accuracy considered in Fig. \ref{nmse_snr}(a), SA1O (SA2O) and SASC rely on fixed-resolution angle grids determined by the numbers of subcarriers and the numbers of  single-carrier scans, respectively,  and the  accuracy is unaffected by the SNR. SA2O outperforms SA1O and matches SASC's performance due to the enhanced diversity gain resulting from the use of  two OFDM symbols.  The ANM method applied in MA1O excels at lower SNR levels, achieving a high accuracy of $0.001{^\circ}$ at $\mathrm{SNR}=-20\textrm{ dB}$.    Compared to MA1O, MASC provides higher precision for  angle estimation,  benefiting from zero IUI and low noise at the narrowband receiver. Then, as depicted in Figs. \ref{nmse_snr}(b) and \ref{nmse_snr}(c),  for velocity and distance estimation,  MA1O  achieves estimation accuracies of  0.1 m/s and  $10^{-3}$ m, respectively,  which are notably higher than those achieved by other schemes.
SA2O's performance  marginally improves with increasing SNR and saturates at  $\mathrm{SNR}=-20\textrm{ dB}$, with accuracy values of 0.2 m/s and 0.5 m for velocity and  distance, respectively. Notably, SA1O, albeit slightly inferior to SA2O, shows similar performance trends.  SASC and MASC  exhibits the worst distance estimation performance, and the accuracy of its velocity estimation is  poor at low SNR, but improves to a level  similar to that of MA1O at high SNR.   In summary, the proposed SA1O  exhibits slightly worse estimation performance than SASC, while the proposed SA2O achieves an estimation performance similar to that of SASC. Meanwhile, the proposed MA1O yields high estimation performance for all considered parameters.

\begin{figure}[htbp]
	\vspace{-0.6cm}
\centering %
\begin{minipage}[t]{0.48\textwidth}%
\centering \includegraphics[width=3.35in]{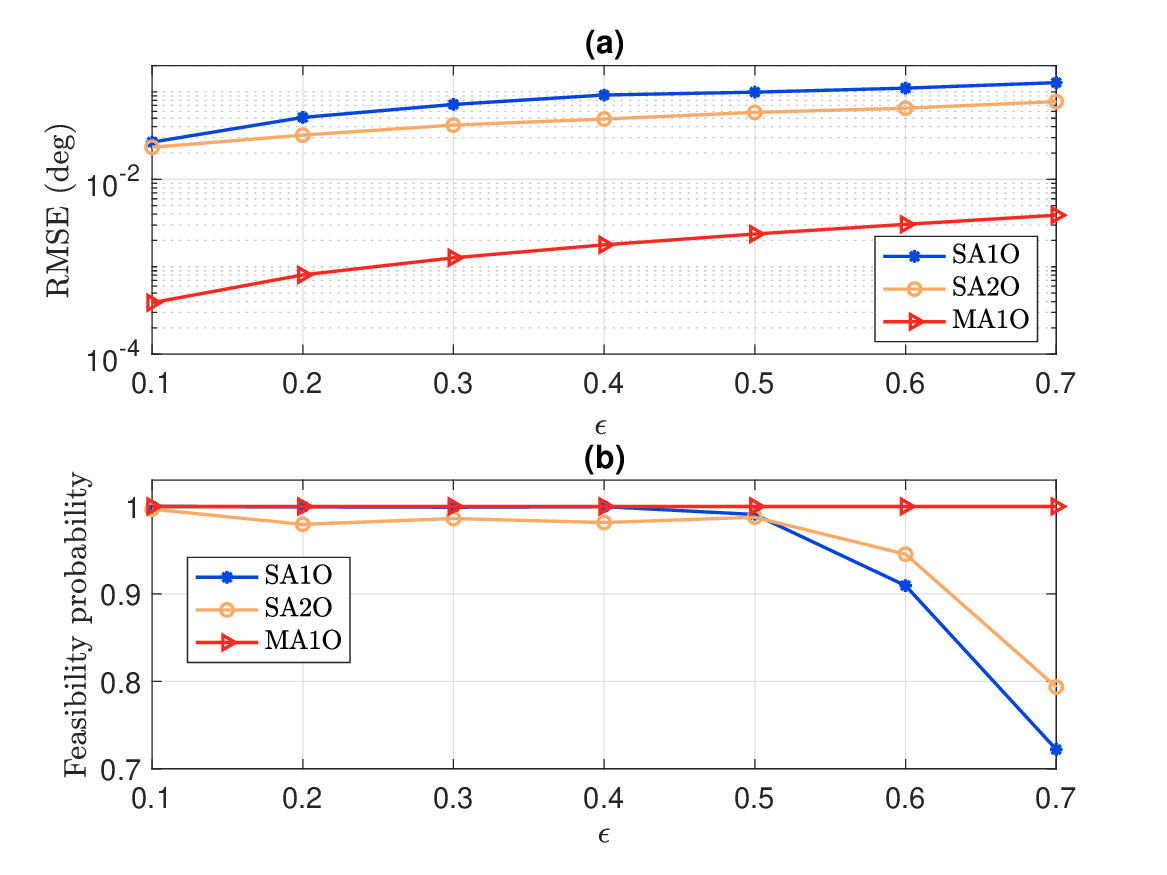} \captionsetup{font={small}}
\caption{Sensing performance versus overlap factor of  
rainbow beams.}
\label{SNR-1-1} %
\end{minipage}\hspace{3mm}
\begin{minipage}[t]{0.48\textwidth}%
\centering \includegraphics[width=3.35in]{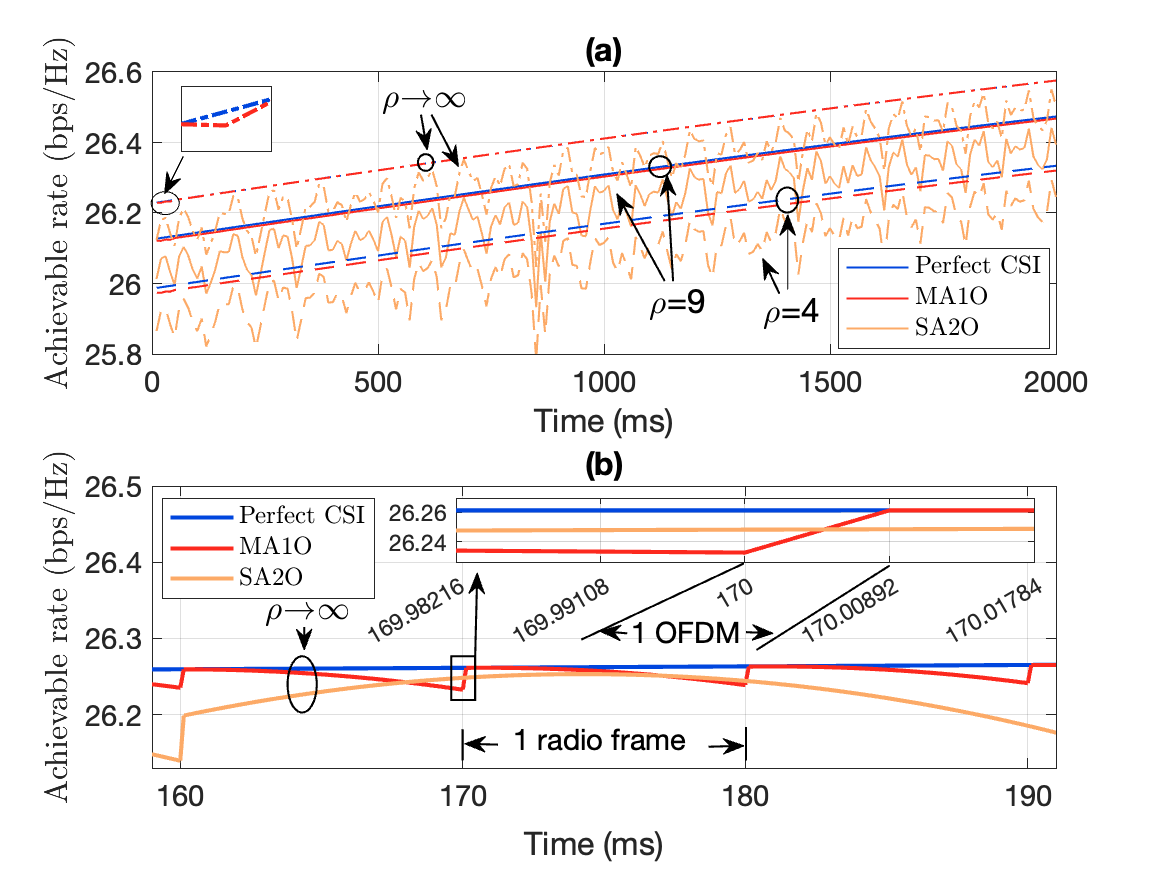} \captionsetup{font={small}}
\caption{Achievable rate versus time for  linear movement trajectory scenario with
velocities of \{45, 50, 55\} m/s. }
\label{rate-city} %
\end{minipage}
\end{figure}

In Fig. \ref{SNR-1-1}, the impact of the beam overlap factor $\epsilon$, used to calculate the minimum number of sensing OFDM subcarriers $M$, on the RMSE of angle  estimation and the feasibility probability is studied, for  $\mathrm{SNR}=10$ dB and $K=1$.  The feasibility probability represents the ratio of angle parameter realizations that result in successful user detection, excluding  false alarms or missed detections, to all angle parameter realizations.  That is,  we categorize the case, where only one angle is detected, as sucessfull in the considered single-user scenario.
The corresponding values of the  $\epsilon$-beamwidth, $\triangle\psi_{\epsilon}$,
and $M$ are provided in Table II.   Fig. \ref{SNR-1-1}(a) shows that  the  RMSE increases  with 
$\epsilon$ for all three schemes, albeit for different reasons. SA1O
and SA2O rely on an angle grid where the resolution is determined by the number of subcarriers, 
$M$. For example, the angle resolution for $\epsilon=0.1$ is $180{^\circ}/1439=0.125{^\circ}$, resulting in an  angle estimation accuracy of $0.02{^\circ}$, c.f. Fig. \ref{SNR-1-1}(a). In contrast, 
$\epsilon=0.7$ yields an angle resolution of $180{^\circ}/452=0.398{^\circ}$,
leading to  angle estimation accuracies of $0.13{^\circ}$ for
SA1O and $0.08{^\circ}$ for SA2O, respectively, c.f. Fig. \ref{SNR-1-1}(a). SA2O benefits
from exploiting two OFDM symbols, which provides an increased number of
measurements for angle estimation, enhancing accuracy compared to
SA1O. MA1O  employs the high-precision ANM method, where angle
resolution is unaffected by the number of  subcarriers. However,  a 
greater beam overlap, as enforced by smaller $\epsilon$,   enhances the number of subcarriers effectively reflected by  the user,  thus providing more precise
measurements for high-accuracy angle estimation. Additionally, Fig. \ref{SNR-1-1}(b) shows  that the feasibility  probabilities of SA1O and SA2O decline when $\epsilon\geq0.5$, indicating more frequent  occurrences of false alarms or missed detections. Therefore, to  ensure reliable detection performance, the beam overlap factor should not exceed $\epsilon=0.5$, corresponding to the 3dB beamwidth.  

\begin{table*}

\centering \centering \caption{Parameters used in Figure \ref{SNR-1-1}.}
\begin{tabular}{|c|c|c|c|c|c|c|c|}
\hline 
\textbf{$\epsilon$}  & 0.1  & 0.2  & 0.3  & 0.4  & 0.5  & 0.6  & 0.7\tabularnewline
\hline 
$\frac{N_{t}}{2\pi}\triangle\psi_{\epsilon}$  & 0.3561  & 0.5151  & 0.6467  & 0.7677  & 0.8860  & 1.0066  & 1.1359\tabularnewline
\hline 
\textbf{$M$}  & 1439  & 995  & 793  & 668  & 579  & 510  & 452\tabularnewline
\hline 
\end{tabular}
\vspace{-0.9cm}
\end{table*}

\subsection{Rainbow Beam User Tracking and Communication}

In this subsection, the user tracking performance is   evaluated based on the achievable rate  for  communication,  leading to optimization problem
\begin{equation}
	\max_{{\bf w}_{k,q}}\frac{1}{Q}\sum_{q=1}^{Q}R_{k,q},\:\textrm{s.t. }|[{\bf w}_{k,q}]_{n}|=1,1\leq n\leq N_{t}.\label{eq:r35-2-1}
\end{equation}
Algorithms for optimization of  ${\bf w}_{k,q}$ based on perfect CSI and based on the estimated  direction parameters of LoS can be found in Appendix \ref{section comu}. According to the 5G frame structure \cite{5gNR-2}, the duration of one radio frame is 10 ms. For a subcarrier spacing of 120 kHz, the duration of an OFDM symbol including the CP is 8.92 $\mu $s, and the duration of a slot containing 14 OFDM symbols is 0.125 ms. In the following,  the   angles of the users are estimated every 10 ms.  For simplicity, the number of subcarriers used for tracking of one user is set to $|\mathcal{M}_{k}|=21$ for all velocities.  The sensing and communication SNRs  are  set to 0 dB,  when  the communication SNR is defined as $\mathrm{SNR}=P_{{\rm c}}\min_{\forall k}\{|\alpha_{k,0}|^{2}\}/\sigma_{\mathrm{c}}^{2}$, where   $P_{{\rm c}}$ and $\sigma_{\mathrm{c}}^{2}$ are the TX power and the noise power at the users, respectively, and $|\alpha_{k,0}|^{2}=\frac{\lambda_{\mathrm{c}}^{2}}{(4\pi)^{2}D_{k}^{2}}$.
The achievable rate obtained based on perfect CSI is regarded as a performance upper bound and denoted by `Perfect CSI'.

First, for linear motion trajectory scenarios, we  study the tracking capabilities of the proposed scheme, focusing on a scenario involving three users moving on a highway. Their velocities are \{45, 50, 55\} m/s, and their initial positions are specified by distances \{370, 448, 600\} m and angles \{$45{^\circ}$, $27{^\circ}$, $40{^\circ}$\}.
In  Fig. \ref{rate-city}(a), we compare  the performances achieved with  precoders designed  based on the estimated LoS angles and the perfect full CSI for various Rice factors.  The achievable rate  is calculated every 10 ms based on the newly updated angles.  Notably, for Rice factor $\rho\rightarrow\infty$,
the communication performance achieved with MA1O-based tracking closely matches for perfect CSI, while SA2O exhibits a noticeable rate loss compared to the perfect CSI case, with the largest  loss in achievable rate reaching approximately 0.15 bps/Hz. For   Rice factors of  $\rho=9$ and $\rho=4$, MA1O-based tracking introduces a small rate loss.  In contrast, the rate loss for  SA2O-based tracking nearly doubles to approximately
0.3 bps/Hz.  In Fig. \ref{rate-city}(b), the achievable rate, calculated for each OFDM symbol, is depicted exemplarily for $\rho\rightarrow\infty$.  During the final few OFDM symbols of each radio frame, the users are tracked, and the updated CSI of LoS path  are employed for communication beamforming design in the next radio frame. It is observed that with  MA1O-based tracking, the initial OFDM symbols within each radio frame achieve excellent transmission rate since the LoS angles are accurately updated. However, the transmission rate gradually decreases for subsequent OFDM symbols within the same radio frame as the CSI information becomes outdated. The achievable rate of SA2O does not increase in every radio frame. We thinks this might be because the rate loss resulting from angle estimation errors could outweigh the impact of outdated angle information in certain radio frames.

\begin{figure}[htbp]
	\vspace{-0.5cm}
\centering %
\begin{minipage}[t]{0.48\textwidth}%
\centering \includegraphics[width=3.35in]{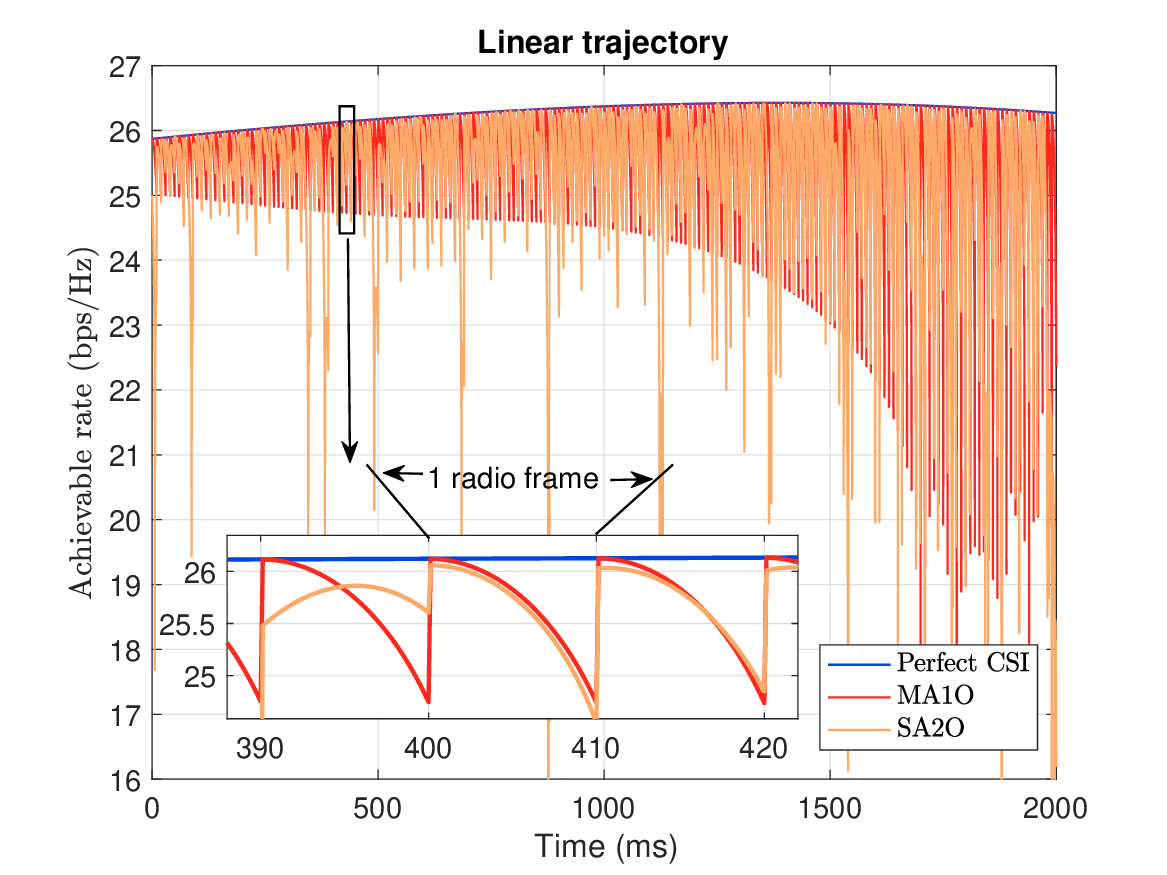}
\captionsetup{font={small}} \caption{Achievable rate versus time for linear 
movement  trajectory scenario with ultra-high velocities 
of \{200, 220, 250\} m/s, where $\rho\rightarrow\infty$.}
\label{rate-plane} %
\end{minipage} \hspace{3mm}
\begin{minipage}[t]{0.48\textwidth}%
\centering \includegraphics[width=3.35in]{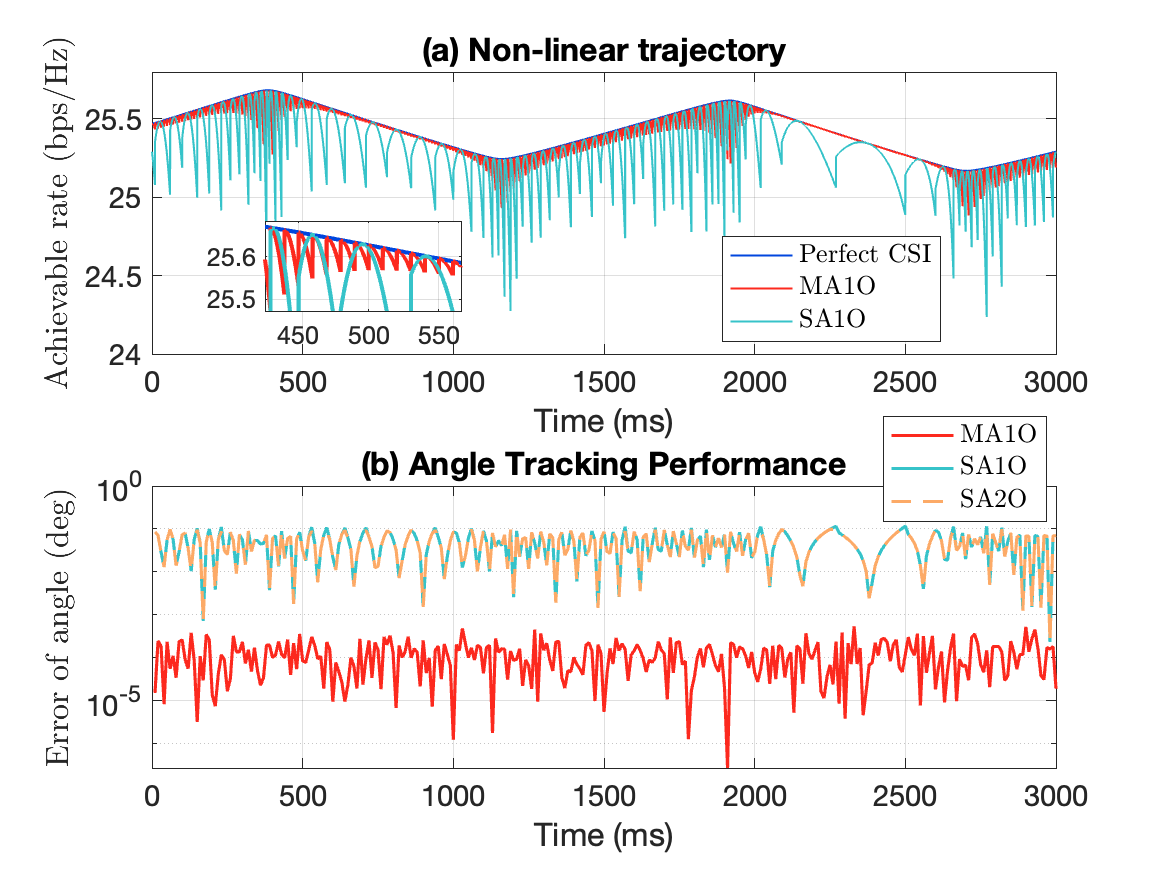} \captionsetup{font={small}}
\caption{Achievable rate versus time for  nonlinear movement trajectory scenario
with velocity  15 m/s, where $\rho\rightarrow\infty$.}
\label{rate-nonlinear} %
\end{minipage}
\end{figure}

Next, in Fig. \ref{rate-plane}, we investigate the achievable rate calculated at every slot   for a  scenario with ultra-high velocities, such as civil
planes with velocities of \{200, 220, 250\} m/s. It is observed that both considered tracking
schemes achieve nearly  precise angle updates within a single radio frame,
facilitating high  communication rates at the start of almost each radio frame. However,
given the exceptionally high user velocities, the angle information
becomes quickly outdated. To ensure dependable communication services
for ultra-high-velocity users, one potential strategy is to increase the tracking frequency or employ beam prediction techniques  \cite{user-tracking}.

\begin{figure}
\end{figure}

Finally,  in Fig. \ref{rate-nonlinear},  we study  the tracking capability
for a nonlinear movement trajectory scenario, where a single user
with a velocity of 15 m/s is moving along a serpentine path, resulting in a non-monotonic  achievable rate as a function of time, 
as shown in Fig. \ref{rate-nonlinear}(a) where  the achievable rate is calculated every slot.  Since SA1O and SA2O achieve similar performance, hence, only the results for SA1O are presented.  MA1O-based tracking consistently provides high  communication rates,  with negligible rate loss for most of the tracking duration.  However, the  rate
loss  becomes more pronounced, when the user transitions through
the trajectory's inflection points, leading to larger angle variations.
In contrast, SA1O-based tracking typically exhibits rate losses
of approximately 0.5 bps/Hz, which indcates suboptimal performance for  nonlinear
trajectory scenarios. Nevertheless,  continuous tracking
of the mobile user is still maintained. From Fig. \ref{rate-nonlinear}(b),  we deduce that the excellent rate
 performance of MA1O can be attributed to its 
ability to provide angle estimates with a precision in the range of
$10^{-5}\sim10^{-4}$, whereas SA1O  yields less precise  angle estimates, typically in the range of $10^{-2}\sim10^{-1}$.

\section{Conclusions}

In this work, we have proposed the application of rainbow beams for ISAC in a mmWave system.  OFDM radar is employed to create frequency-dependent rainbow beams based on one  OFDM symbol, covering the entire angular space for fast initial access. For low-velocity users,
a resource-efficient scheme  based on a  single-antenna radar receiver and one set of rainbow brams  was proposed for estimation of the user parameters.  To address the Doppler ambiguity caused by   high-velocity users,  two enhanced estimation schemes exploiting respectively  two sets of rainbow beams   and a  multi-antenna radar receiver have been introduced.  Furthermore,   a  user tracking scheme avoiding switching of PSs and TTDs  has been proposed by selectively manipulating a subset of subcarriers  from the radar rainbow beam, reducing the demands on control circuitry and minimizing time overhead. Our simulation results indicate that a 3 dB-beamwidth overlap is sufficient to design effective rainbow beams for sensing with a feasibility probability approaching 1. Moreover, the proposed estimation scheme based on a multi-antenna radar receiver yields high-precision estimates for the angles, distances, and velocities of  the users at  medium and high SNRs, realizing excellent mobile data rate performance approaching that for perfect CSI.

\appendices{}

\section{Optimization of Achievable Rate for Communication}

\label{section comu}

We consider a simple communication scenario to  demonstrate the mitigation of the beam squint for communication and the utilization of the user parameters obtained during sensing in
the proposed ISAC framework.

Denote by $\mathbf{t}_{k,q}^{{\rm com}}=[e^{-j2\pi f_{k,q}^{\rm com}\kappa_{k,1}},e^{-j2\pi f_{k,q}^{\rm com}\kappa_{k,2}},\ldots,e^{-j2\pi f_{k,q}^{\rm com}\kappa_{k,N_{t}}}]^{\mathrm{T}}$
the TTD vector for the $k$-th user on the $q$-th subcarrier. Define
delay vector $\boldsymbol{\kappa}_{k}=[\kappa_{k,1},...,\kappa_{k,N_{t}}]^{{\rm T}}$,
where each element is limited to $\kappa_{k,n}\in[0,1]/{f_{k,q}^{\rm com}},1\leq q\leq Q$,
leading to the constraint $\kappa_{k,n}\in[0,1]/{f_{k,Q}^{\rm com}}$.
The aim of designing precoding vector $\mathbf{w}_{k,q}={\bf w}_{\mathrm{PS},k}\odot\mathbf{t}_{k,q}^{{\rm com}}$
is to maximize the average rate over all the subcarriers allocated to the  $k$-th
user,  leading to optimization problem
\begin{equation}
	\max_{{\bf w}_{\mathrm{PS},k},\boldsymbol{\kappa}_{k}}\frac{1}{Q}\sum_{q=1}^{Q}R_{k,q},\:\textrm{s.t. }|[{\bf w}_{\mathrm{PS},k}]_{n}|=1,\kappa_{k,n}\in[0,1]/{f_{k,Q}^{\rm com}},1\leq n\leq N_{t}.\label{eq:r35-2-1}
\end{equation}

Using Jensen's inequality, an upper-bound for the objective function in (\ref{eq:r35-2-1}) is derived as $\frac{1}{Q}\sum_{q=1}^{Q}R_{k,q}\leq\log_{2}\left(\frac{1}{Q}\sum_{q=1}^{Q}\left(1+\frac{P_{{\rm c},k}}{\sigma_{{\rm c},k}^{2}}|\mathbf{h}_{k,q}^{\mathrm{H}}{\bf w}_{k,q}|^{2}\right)\right)$ \cite[Equ. (17)]{OFDM-mmwave}.
Since the main focus of this paper is the proposed rainbow beam based training and tracking, we simplify Problem (\ref{eq:r35-2-1}) by optimizing the upper-bound on the average rate.  This lead to the following optimization problem, 
\begin{equation}
	\max_{{\bf w}_{\mathrm{PS},k},\boldsymbol{\kappa}_{k}}\sum_{q=1}^{Q}|\mathbf{h}_{k,q}^{\mathrm{H}}{\bf w}_{k,q}|^{2},\:\textrm{s.t. }|[{\bf w}_{\mathrm{PS},k}]_{n}|=1,\kappa_{k,n}\in[0,1]/{f_{k,Q}^{\rm com}},1\leq n\leq N_{t},\label{eq:r35-3}
\end{equation}
where $|\mathbf{h}_{k,q}^{\mathrm{H}}{\bf w}_{k,q}|^{2}={\bf w}_{\mathrm{PS},k}^{{\rm H}}{\rm diag}((\mathbf{t}_{k,q}^{{\rm com}})^{{\rm H}})\mathbf{h}_{k,q}\mathbf{h}_{k,q}^{\mathrm{H}}{\rm diag}(\mathbf{t}_{k,q}^{{\rm com}}){\bf w}_{\mathrm{PS},k}$.

We firsty assume that the  full CSI for $\text{\ensuremath{\mathbf{h}}}_{k,q}$ is available. Given delay vector $\boldsymbol{\kappa}_{k}$, the optimal ${\bf w}_{\mathrm{PS},k}$
for  Problem (\ref{eq:r35-3}) is given by ${\bf w}_{\mathrm{PS},k}^{\mathrm{opt}}=\exp\left\{ j\angle\mathbf{u}_{k}\right\} $,
where $\mathbf{u}_{k}$ is the eigenvector corresponding to the largest
eigenvalue of matrix $\sum_{q=1}^{Q}{\rm diag}((\mathbf{t}_{k,q}^{{\rm com}})^{{\rm H}})\mathbf{h}_{k,q}\mathbf{h}_{k,q}^{\mathrm{H}}{\rm diag}(\mathbf{t}_{k,q}^{{\rm com}})$
and $\exp\left\{ j\angle\left(\cdot\right)\right\} $ is an element-wise
operation.  Given ${\bf w}_{\mathrm{PS},k}$,  the optimal $\mathbf{t}_{k,q}^{{\rm com}}$ is obtained as  $(\mathbf{t}_{k,q}^{{\rm com}})^{\rm opt}=\exp\left\{ j\angle\left({\rm diag}({\bf w}_{\mathrm{PS},k}^{{\rm H}})\mathbf{h}_{k,q}\right)\right\} $ if we ignore the dependence of  $\boldsymbol{\kappa}_{k}$ for the moment. Then, the optimal delay $\boldsymbol{\kappa}_{k}$
should satisfy
\begin{equation}
	\min_{\boldsymbol{\kappa}_{k}}\sum_{q=1}^{Q}\left\Vert \boldsymbol{\kappa}_{k}+\frac{1}{2\pi f_{k,q}^{\rm com}}\angle\left((\mathbf{t}_{k,q}^{{\rm com}})^{\rm opt}\right)\right\Vert _{2}^{2},\:\textrm{s.t. }\kappa_{n}\in[0,1]/{f_{k,Q}^{\rm com}},1\leq n\leq N_{t},\label{eq:r35-30}
\end{equation}
and is given by $\boldsymbol{\kappa}_{k}^{{\rm opt}}=-\frac{1}{Q}\sum_{q=1}^{Q}\left(\frac{1}{2\pi f_{k,q}^{\rm com}}\angle\left({\rm diag}({\bf w}_{\mathrm{PS},k}^{{\rm H}})\mathbf{h}_{k,q}\right)\right)$.
Finally, ${\bf w}_{\mathrm{PS},k}^{\mathrm{opt}}$ and $\boldsymbol{\kappa}_{k}^{{\rm opt}}$
are updated iteratively.

However, when only the  CSI of the LoS path is available in terms of AoDs and propagation delays, as obtained via user training and tracking, the suboptimal
problem corresponding to Problem (\ref{eq:r35-3}) is given by 
\begin{align}
	\max_{{\bf w}_{\mathrm{PS},k},\boldsymbol{\kappa}_{k}} & \sum_{q=1}^{Q}{\bf w}_{k,q}^{\mathrm{H}}\mathbf{a}_{N_{t}}\left(\theta_{k},f_{{\rm c}}+f_{k,q}^{\rm com}\right)\mathbf{a}_{N_{t}}^{\mathrm{H}}\left(\theta_{k},f_{{\rm c}}+f_{k,q}^{\rm com}\right){\bf w}_{k,q},\nonumber \\
	\textrm{s.t. } & |[{\bf w}_{\mathrm{PS},k}]_{n}|=1,\kappa_{n}\in[0,1]/{f_{k,Q}^{\rm com}},1\leq n\leq N_{t}.\label{eq:r35-4}
\end{align}
According to Section (\ref{beamscom}), the optimal delay vector is given by
$\boldsymbol{\kappa}_{k}^{{\rm opt}}=\frac{d}{c}\cos\theta_{k}[0,1,...,N_{t}-1]^{{\rm T}}$,
and the optimal phase shift vector is given by ${\bf w}_{\mathrm{PS},k}^{\mathrm{opt}}=\mathbf{a}_{N_{t}}\left(\theta_{k},f_{{\rm c}}\right)$.

 \bibliographystyle{IEEEtran}
\bibliography{bibfile}

\end{document}